\documentclass[fleqn,10pt]{olplainarticle}
\usepackage{times}
\usepackage{amsmath}
\usepackage{xcolor}
\usepackage[numbers,square,sort, compress]{natbib}
\usepackage{float}% use to force figures in place

\usepackage{fancyhdr,ifoddpage} % for running title
\fancypagestyle{runningheads}{%
	\fancyhf{}% Clear header/footer
	\fancyhead[L]{% Left header
		\checkoddpage% Update \ifoddpage
		\ifoddpage
		N.~Margaritella, V.~Inácio, R.~King% Odd pages
		\else
		Bayesian fPCA with multilevel partition priors for neuroscientific studies% Even pages
		\fi
	}%
	\fancyhead[R]{\thepage}% Right header
}

\title{A Bayesian functional model with multilevel partition priors for group studies in neuroscience}

\author[1$^*$]{Nicol\`{o} Margaritella} 
\author[2]{Vanda In\'{a}cio}
\author[3]{Ruth King}
\affil[1$^*$]{School of Mathematics and Statistics, University of St Andrews, St Andrews, KY16 9SS, U.K.\\ E-mail:nm256@st-andrews.ac.uk. https://orcid.org/0000-0003-3776-2773}
\affil[2,3]{School of Mathematics and Maxwell Institute for Mathematical Sciences, University of Edinburgh, Edinburgh, U.K. }


\addinfo{Word count: 7597; Tables: 1; Figures: 5; Supplementary materials included in a separate file.}

\begin{abstract}
	The statistical analysis of group studies in neuroscience is particularly challenging due to the complex spatio-temporal nature of the data, its multiple levels and the inter-individual variability in brain responses.  In this respect, traditional ANOVA-based studies and linear mixed effects models typically provide only limited exploration of the dynamic of the group brain activity and variability of the individual responses potentially leading to overly simplistic conclusions and/or missing more intricate patterns. In this study we propose a novel Bayesian model-based clustering method for functional data to simultaneously assess group effects and individual deviations over the most important temporal features in the data. To this aim, we develop an innovative multilevel partition prior to model the functional scores of a functional Principal Components decomposition of neuroscientific recordings; this approach returns a thorough exploration of group differences and individual deviations without compromising on the spatio-temporal nature of the data. By means of a simulation study we demonstrate that the proposed model returns correct classification in different clustering scenarios under low and high noise levels in the data. Finally we consider a case study using Electroencephalogram data recorded during an object recognition task where our approach provides new insights into the underlying brain mechanisms generating the data and their variability.
\end{abstract}

\begin{document}
%	\footnote{abcd}
	
	\flushbottom
	\maketitle
	\thispagestyle{empty}

%\footnote*{\textbf{Corresponding Author: Nicolò Margaritella}, School of Mathematics and Statistics, University of St Andrews, St Andrews, KY16 9SS, U.K. E-mail: nm256@st-andrews.ac.uk. https://orcid.org/}

%% maketitle must follow the abstract.

%% If there is not enough space inside the running head
%% for all authors including the title you may provide
%% the leftmark in one of the following three forms:

%% \renewcommand{\leftmark}
%% {First Author: A Short Title}

%% \renewcommand{\leftmark}
%% {First Author and Second Author: A Short Title}

%% \renewcommand{\leftmark}
%% {First Author et al.: A Short Title}

%% \tableofcontents  % Produces the table of contents.

\section{Introduction}
\label{s:intro}
Group comparison for neuroscientific data involves comparing sets of spatio-temporal recordings such as those obtained using electroencephalography (EEG), magnetoencephalography or functional Magnetic Resonance Imaging (fMRI) under different conditions. The spatio-temporal nature of the data combined with its multiple levels structure (recording, subject, group) and the inter-individual heterogeneity in brain responses, make the analysis of these data an open challenge in neuroscience.

ANOVA-based analyses have been the standard approach for task-based experiments in neuroscience \cite{alday2017electrophysiology}. Although well suited for several experimental designs, ANOVA methods in neuroscience have limited the exploration of complex heterogeneous spatio-temporal processes to a simple binary decision (accept/reject) over aggregated measures of the brain signals (e.g. latency, amplitude, power in clinically relevant frequency bands). More recently, the use of complex dynamic stimuli in cognitive neuroscience has driven the choice of statistical methods toward linear mixed-effect models \cite{yu2022beyond,alday2017electrophysiology,hasson2012future}. Linear mixed effect models allow more flexible designs and also account for additional variation at multiple levels. Despite the increased flexibility, in general, standard mixed effect models are insufficient for the study of neuroscientific recordings arising from group studies as they are not able to capture how different sources of heterogeneity, such as group effects and individual deviations, vary in time. Individual differences might arise for several reasons, e.g. they could be linked to subject-specific conditions such as tiredness and caffeine intake; the experimental design, e.g. lengthy experiments with repeated tasks are linked to stimulus habituation and fatigue; and/or expressions of different brain strategies within a target population.

Most types of brain recordings can typically be well represented by noisy realisations of smooth time processes, hence Functional Data Analysis (FDA) is an ideal framework for the modelling of neuroscientific data. The FDA literature is vast, and we recommend  \citet{ramsay2005functional} for an introduction to the topic and \citet{gertheiss2023functional} for a recent review.
Among FDA methods, functional Principal Component Analysis (fPCA) has been successfully employed in neuroscientific studies since the early 2000s. For example, \citet{viviani2005functional} used fPCA for the analysis of fMRI data showing that fPCA is more effective in recovering the signal of interest than standard PCA. %\citet{di2009multilevel} proposed a multilevel functional PCA model to extract intra- and inter-subject geometric components of multilevel functional data which was successfully employed to identify and quantify associations between EEG activity during sleep and adverse cardiovascular outcomes.
\citet{hasenstab2017multi} proposed a multi-dimensional version of fPCA for the analysis of repeated evoked potentials in EEG data showing how the new methodology provided further insights into the learning patterns of children with Autism Spectrum Disorder. The use of fPCA to explore complex spatio-temporal patterns is still an ongoing area of research \citep{wu2022clustering}.

Functional clustering is another promising research area for the analysis of neuroscientific recordings \citep{zhang2023review}. Clustering of brain recordings could help explore how brain activity is organised. Early methods for clustering functions focused on clustering time series through the clustering of the relative coefficients in a basis expansion representation \citep{angelini2012clustering,ray2006functional,zhou2006bayesian,james2003clustering}. 
More recently, the exploration of brain activity from a dynamical perspective and the understanding of  inter-individual variability in brain functions have attracted wide interest in the neuroimaging literature \citep{gonzalez2018task,warnick2018bayesian,hutchison2013dynamic,seghier2018interpreting,li2017linking,zhang2016spatiotemporal}. %In this context, neuroscientists are interested in unveiling how functional connections between brain areas \textit{evolve in time} under different conditions or in different groups and how they vary in a target population.

From a functional clustering perspective, the study of dynamic brain patterns requires clustering of local features of the curves.
The idea of local clustering was firstly introduced by \citet{dunson2009nonparametric}, using partition priors, and \citet{petrone2009hybrid}  by means of canonical curves, both in an attempt to capture local variations in the curves. \citet{suarez2016bayesian} and \citet{margaritella2021parameter} used Dirichelet Process (DP) priors to cluster wavelet coefficients and functional Principal Component scores, respectively; the latter showed how combining fPCA and Bayesian model-based clustering can produce an in-depth exploration of spatio-temporal patterns in neuroscientific recordings. 

In this paper we introduce an innovative approach for group studies in neuroscience which explores differences as well as inter-individual variability \textit{dynamically}, as they evolve in time. We propose a novel comprehensive model-based approach for the exploration of dynamic brain activity patterns in the context of group studies, namely Bayesian fPCA with multilevel partition priors, with the specific aims of (a) identifying what is the impact of different experimental conditions on brain activity at different time windows; and (b) detecting if and how these time-varying effects differ among individuals in the same group. By achieving these aims we are able to provide neuroscientists with a better and more complete understanding of underlying brain mechanisms and their variability in a target population. 

At the core of our method we develop a novel partition prior, called multilevel partition prior, which allows for local clustering at different levels in the data, namely between all individuals (common cluster), between subjects of the same group (group-specific clusters) and within individuals (subject-specific clusters). The first two levels allow for the identification of different local group effects while the latter captures any possible individual deviations from the group trends (outlying features). Our proposed approach directly models the both group differences and individual-specific deviation from group trends, accounting for both group and individual level variability within a consistent and structured framework that suits well the context of group studies.  The \textit{local} nature of the proposed prior means that a measure of the strength of local group effects and local inter-subject heterogeneity is obtained a posteriori. In addition, relevant local features can involve potentially different subgroups of the original study cohort, returning intricate results with the potential to provide novel insights on the dynamic brain activity in groups under different experimental conditions and of the associated inter-individual heterogeneity.

The paper is organised as follows: in Section 2 we present the features of the proposed model; in Section 3 we show results of a simulation study to test the model performance under low and high noise levels in the data; results of a case study employing EEG recordings are reported in Section 4; conclusions and further directions are discussed in Section 5. 
	
	\section{Bayesian functional PCA with multilevel partition priors}
	\label{Chap2}
	In this section we present the structure of the proposed model and the key features of this approach.
	\subsection{Hierarchical structure}
	\label{chap21}
	The following hierarchical structure defines the probability distributions for the data and all the parameters in the model. Each hierarchical level is introduced and commented separately starting from the data level.\vspace*{6pt}\\
	%\vspace*{5pt}
	\noindent\textit{Level 1:} we assume that data $ Y_{ui}(t) $ from $ i=1,\dots,n $ recording locations and $ t=1,\dots,T $ time points are collected from $ u=1,\dots,U $ units. Further, we assume that $ u=1,\dots,U_{A} $ are units from group A and $ u=U_{A}+1,\dots,U $ are units from group B. Groups could represent, for example, patients and controls; or individuals undertaking two different tasks. For simplicity, we will assume that units identify subjects from either group A or B for the rest of  this Section.
	%Further we assume that $ u=1,\dots,U_{A} $ are subjects from group A and $ u=U_{A}+1,\dots,U $ are subjects from group B. Groups could represent, for example, patients and controls; or individuals undertaking two different tasks.
	Following standard functional data analysis (FDA) assumptions \citep{ramsay2005functional}, we define the distribution of the centred data $ \widetilde{Y}_{ui}(t) $, given the parameters $ \xi_{uik} $ of the underlying smooth functions $ X_{ui}(t) $ and the noise term $ \tau $, to be:
	\begin{eqnarray}\label{eq1}
		\tilde{\textbf{Y}}_{ui}|\textbf{X}_{ui},\tau & \sim & \text{N}_{\text{T}}(\textbf{X}_{ui},\tau^{-1}\textbf{I}),\\\nonumber
		\textbf{X}_{ui} & = & \sum_{k=1}^{K}\xi_{uik}\boldsymbol{\nu}_{k},
	\end{eqnarray}
	where $ \tilde{\textbf{Y}}_{ui} $, $ \textbf{X}_{ui} $, and the eigenfunctions $ \boldsymbol{\nu}_{k} $ are \textit{T}-dimensional vectors and $ \text{N}_{\text{T}}(\textbf{X}_{ui},\tau^{-1}\textbf{I})  $ denotes the probability density function of a multivariate Gaussian random variable with mean $ \textbf{X}_{ui} $ and variance-covariance matrix $ \tau^{-1}\textbf{I} $, such that $ \textbf{I} $ denotes a $ T\times T $ identity matrix. The parameters $\xi_{uik}$ are the functional Principal Component scores (fPC scores) and the focus of our analysis. We use a set of common bases obtained from the eigenanalysis of the total covariance function, $\text{Cov}\{X_{ui}(s),X_{ui}(t)\}$, which let us explore variation at different levels through clustering of the fPC scores. The principal component bases have the advantage of being the most efficient bases of size $ K $ and, most importantly, they offer a meaningful interpretation in terms of explained variance \citep{ramsay2005functional}. The number of bases, $K$, is selected to include the most important bases up to a pre-defined amount of total variation (examples are provided in both the simulation and real data analyses). This selection is consistent with the multilevel structure of the data; in fact, leading eigendimensions will most likely capture group behaviours while trailing eigendimensions will increasingly show subject-specific features.
	We assume constant noise for simplicity, noting that including an additional model for the noise will increase computational expense with expected limited benefits for the analysis \citep{ramsay2005functional}.\vspace*{6pt}\\
	%
	% LEVEL 2 - PRIORS
	\textit{Level 2:} to indicate the latent classes and the relative parameters, we introduce a classification variable $ z_{uik} $ that identifies which latent class $ j $ is associated with parameter $ \xi_{uik} $. For each $u,i,k$ with $k=1,\dots,K$, the prior conditional distribution of $ \xi_{uik} $, given the parameters of underlying clusters $ \big[(\mu_{uk1},s_{uk1}),\dots,(\mu_{ukJ},s_{ukJ})\big] $ and the classification variable $ z_{uik} $, is given by
	\begin{eqnarray}\label{eq2}
		\xi_{uik}|z_{uik},\mu_{ukj},s_{ukj}&\sim &\text{N}\big(\mu_{u\,k,\, j=z_{uik}},s^{-1}_{u\,k,\, j=z_{uik}}\big),
	\end{eqnarray}
	where $ \mu_{u\,k,\, j=z_{uik}} $ and $ s_{u\, k,\, j=z_{uik}} $ are the corresponding mean and precision for the $ j $-th cluster of the $ u $-th subject in the $ k $-th eigendimension, respectively. It is worth noting in Equation~(\ref{eq2}) that $ z_{uik}=z_{u'ik} $ and $ z_{uik}=z_{ui'k} $ do not imply $ \xi_{uik}=\xi_{u'ik} $ and $ \xi_{uik}=\xi_{ui'k} $ respectively, but only that these fPC scores share the same cluster means and variances for a given eigendimension $ k $. Therefore, our approach allows for both within-subject and inter-subject variability to be considered in our model structure through variability in the fPC scores. 
	We employ a $ J- $dimensional mixture of Gaussian distributions, independently for each retained eigendimension $ k=1,\dots,K $ to allow for different partitions in each mode of variation.
	A key role in our model is played by the classification variable $ z_{uik}=j $ with $j=1,\dots,J$ defined as:
	\begin{eqnarray}\label{eq3}
		z_{uik}& = & c^{[1]}_{g_{uk}}+c^{[2]}_{g_{uk}}D_{uk}+c^{[3]}_{g_{uk}}\eta_{uik}\,.
	\end{eqnarray}
	The indicator vector $\boldsymbol{c}_{g_{uk}}=\left(c^{[1]}_{g_{uk}},c^{[2]}_{g_{uk}},c^{[3]}_{g_{uk}}\right)\in \{(1,0,0),(0,1,0),(0,0,1)\} $, controlled by the random variable $g_{uk}$, is defined such that $ c^{[1]}_{g_{uk}}$ corresponds to the label $j=1$ and denotes membership to a \textit{common cluster}; $ c^{[2]}_{g_{uk}}$ corresponds to labels $j=2,3,$  which denote membership to a \textit{group-specific cluster} (either group A or B, according to a fix indicator $ D_{uk}\in \{2,3\} $); and $ c^{[3]}_{g_{uk}}$ corresponds to labels $ j=4,\dots,J $  which denote membership to a \textit{subject-specific cluster}, and are allocated according to the variable $ \eta_{uik} $. We refer to this partition of labels $\{\{1\},\{2,3
	\},\{4,\dots,J\}\}$ as the \textit{subject-level partition}. The variable $\eta_{uik}$ regulates a second level of clustering called \textit{recording-level partition} to explore clustering of subject-specific recording locations $i=1,\dots,N$ in order to identify possible deviations from the relative group behaviour. This paradigm can be extended to include more than two groups by allocating additional labels to the group-specific clusters and shifting the labels for the subject-specific clusters accordingly.
	
	To clarify the notation, consider the following example for a specific eigendimension $k$. When two or more subjects from groups A and B share similar patterns in $k$, all their fPC scores receive label 1 and they are clustered together in the common cluster. On the other hand, when two or more subjects from the same group share patterns not found in the other group, their relative fPC scores receive either label 2 (if in group A) or 3 (if in group B) and are clustered together in the relative group-specific cluster. Finally, if a subject shows patterns that are different from those captured by the global clusters (common and group), a second level of clustering, the \textit{recording-level partition} is activated to assign the subject's scores to subject-specific clusters (with labels ${4,\dots,J}$) according to the variable  $ \eta_{uik} $. It is worth noting that this approach, in principle, can model a variety of scenarios ranging from a single common cluster to $U$ subject-specific clusters; hence, it allows an intricate network of knowledge exchange between fPC scores providing a flexible overall global structure of dependence among locations and subjects. \vspace*{6pt}\\
	%
	% LEVEL 3 - HYPERPRIORS FOR THE CLUSTERS
	\textit{Level 3:} prior distributions for $ \big[(\mu_{uk1},s_{uk1}),\dots,(\mu_{ukJ},s_{ukJ})\big]  $, $ g_{uk} $ and $ \eta_{uik} $ and the hyperparameters $ \phi_{ukj},\,\gamma_{ukj},\, (\omega_{1Dk},\omega_{2Dk}, \omega_{3Dk}) $ and $ (p_{1Dk},\dots,p_{JDk}) $, are given by
	\begin{eqnarray}\label{eq4}
		\eta_{uik}|p_{1Dk},\dots,p_{JDk}&\sim& f_{\text{C}}\big(p_{1Dk},\dots,p_{JDk}\big),\\\nonumber
		g_{uk}|\omega_{1k},\omega_{2k}, \omega_{3k}&\sim& f_{\text{C}}(\omega_{1k},\omega_{2k}, \omega_{3k}),
	\end{eqnarray}
	\begin{alignat*}{4}
		\mu_{ukj}=&\mu_{k 1}\quad	&\mu_{k1}&\sim\text{N}(0,h^{-1}_{k1})	&\quad\text{for }j&=1,\\\nonumber
		\quad s^{-1/2}_{ukj}=&s^{-1/2}_{k1}\quad	&	s^{-1/2}_{k1}&\sim \text{U}(0,\gamma_{k1})	&\quad\text{for }j&=1,\\[0.2cm]\nonumber
		\quad\mu_{ukj}=&\mu_{kD}\quad	&\mu_{kD}&\sim\text{N}(\phi_{kD},h^{-1}_{kD})	&\quad\text{for }j&=2,3,\\\nonumber
		\quad s^{-1/2}_{ukj}=&s^{-1/2}_{kD}\quad	& s^{-1/2}_{kD}&\sim \text{U}(0,\gamma_{kD})	&\quad\text{for }j&=2,3,\\[0.2cm]\nonumber	
		\quad&\quad	&\mu_{ukj}&\sim \text{N}(0,h^{-1}_{ukj})	&\quad\text{for }j&=4,\dots,J,\\\nonumber
		\quad&\quad	&s^{-1/2}_{ukj}&\sim \text{U}(0,\gamma_{ukj})	&\quad\text{for }j&=4,\dots,J,\nonumber
	\end{alignat*}
	where $f_{\text{C}}()$ and $\text{U}()$ denote the probability mass function of a categorical random variable and the probability density function of a uniform random variable, respectively. Parameters $ (\omega_{1k},\omega_{2k},\omega_{3k}) $ control the subject-level partition probabilities in each eigendimension (note that $\sum_{k=1}^{K}\omega_{k}=1$). For example, if $ \omega_{11}\approx 1 $,
	all subjects' fPC scores in eigendimension $ 1 $ would be assigned to the common cluster with mean $ \mu_{11} $ and precision $ s_{11} $ (or, equivalently, standard deviation $s^{-1/2}_{11}$). In the opposite scenario, $ \omega_{31}\approx 1 $ would assign all subjects in eigendimension $ 1 $  to subject-specific clusters. When a subject is allocated to a subject-specific cluster, the recording-level partition probabilities $ (p_{1Dk},\dots,p_{JDk}) $ govern the allocation of each subject's fPC scores to a potentially different number of clusters, labelled $ j\in\{4,\dots,J\}$, and with mean $ \mu_{u1j} $, precision $ s_{u1j} $ and cluster size controlled by $ p_{jD1} $. By making these probabilities dependent on the groups $D$, we account for group-specific differences in the number and size of the subject-specific clusters.
	Hyperparameters $h, \gamma, \phi$ can be centred around functions of the empirical fPC score estimates \citep{richardson1997bayesian}. In practice, using the empirical means and variances of the fPC scores for every group $D$ and dimension $ k $ appeared to work well in both the simulation and case study. A detailed summary of the different settings used is reported in Supplementary Materials A.\vspace*{6pt}\\
	%
	% LEVEL 4 - HYPERPRIORS FOR THE LEVELS
	\textit{Level 4:} prior distributions for $ (\omega_{1k},\omega_{2k},\omega_{3k}) $ and $ (p_{1Dk},\dots,p_{Dk}) $ are given by
	\begin{eqnarray}\label{eq5}
		p^{*}_{1Dk}|\alpha_{k}&\sim&\text{Beta}(1,\alpha_{k}),\\\nonumber
		p_{1Dk}=\dfrac{p^{*}_{1Dk}}{\sum_{j=1}^{J}p^{*}_{jDk}},&&\quad
		p_{jDk}=\dfrac{p^{*}_{jDk}\prod_{l<j}(1-p^{*}_{lDk})}{\sum_{j=1}^{J}p^{*}_{jDk}},\\\nonumber
		\boldsymbol{\omega}_{k}&\sim&\text{Dir}\bigg(\delta_{1},\delta_{2},\delta_{3}\bigg),
	\end{eqnarray}
	where $\text{Dir}()$ represents the Dirichlet distribution. Setting $\delta_{1}=\delta_{2}=\delta_{3}$ produces no prior advantage between the three cluster types (common, group-specific and subject-specific) but, in practice, setting  $\delta_{1}=\delta_{2}<\delta_{3}$ offers a prior better aligned to the usual prior beliefs of a group study (equal prior probability of common vs group effects and low chance of subject-specific deviations).
	The parameters $\boldsymbol{p}_{Dk}$ follow the stick-breaking construction \citep{sethuraman1994constructive} with parameter $ \alpha_{k} $ modelling the prior belief over the mixing proportions $ p_{1Dk},\dots,p_{JDk} $. The dispersion parameter $ \alpha $ is usually fixed or modelled with a prior distribution. Both the expected number of clusters and their variance can be obtained for fixed $\alpha_{k}$ and number of recording sites $n$ or by averaging out the prior distribution for $\alpha_{k}$ \citep{escobar1994estimating,jara2007dirichlet}. We used $\alpha_{k}=1$ as it represents our prior belief that a limited number of clusters is expected to arise in a subject's recording and we let this expectation gradually grow with the number of recoding sites $n$. Details of the choice of $\delta$s and $\alpha$  are reported in Supplementary Materials A. In practice, a simulation study should be conducted to assess the sensitivity of the posterior estimates to the priors specified.
	
\begin{figure}
	\begin{center}
		\includegraphics[,width=0.6\linewidth]{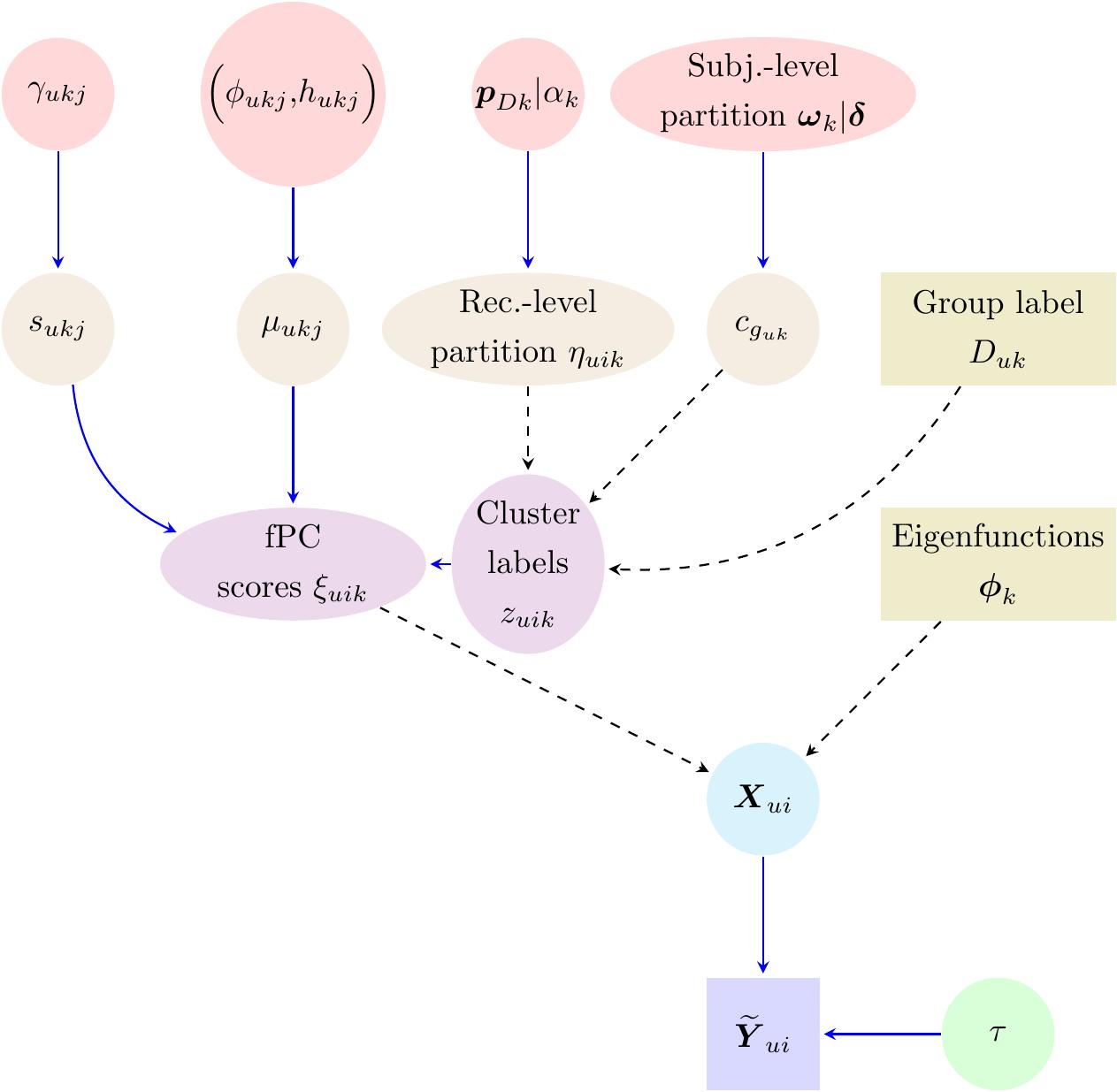}
	\end{center}
	\caption{The structure of our proposed model shown with a Direct Acyclic Graph (DAG). Squares= observed random variables; circles=  unobserved random variables; dashed= deterministic relationship; solid= stochastic relationship. Rec.-level and Subj.-level are abbreviations for Recording-level and Subject-level partitions, respectively.
		\label{Fig1}}
\end{figure}

	The complete model structure can be displayed with a direct acyclic graph (DAG, Figure~\ref{Fig1}).
	The joint posterior distribution of all parameters  is given by
	\begin{align}
		p(\boldsymbol{\xi},\boldsymbol{g}, \boldsymbol{\eta},\boldsymbol{\mu}, \boldsymbol{s},\boldsymbol{\omega},\tau|\mathbf{\tilde{Y}}, \boldsymbol{\nu},\boldsymbol{\phi},\boldsymbol{h}, \boldsymbol{\gamma},\boldsymbol{\delta}, \boldsymbol{\alpha})&\propto f(\mathbf{\tilde{Y}}|\boldsymbol{\xi},\tau,\boldsymbol{\nu})p(\boldsymbol{\xi},\boldsymbol{g}, \boldsymbol{\eta},\boldsymbol{\mu}, \boldsymbol{s}, \boldsymbol{\omega},\tau|\boldsymbol{\phi},\boldsymbol{h}, \boldsymbol{\gamma},\boldsymbol{\delta}, \boldsymbol{\alpha})\\ 
		&=f(\mathbf{\tilde{Y}}|\boldsymbol{\xi},\tau,\boldsymbol{\nu})p(\boldsymbol{\xi}|\boldsymbol{g}, \boldsymbol{\eta},\boldsymbol{\mu}, \boldsymbol{s})p(\boldsymbol{g}|\boldsymbol{\omega})p(\boldsymbol{\eta}|\boldsymbol{\alpha}) \nonumber\\&\quad p(\boldsymbol{\omega}|\delta)p(\boldsymbol{\mu}|\boldsymbol{\phi},\boldsymbol{h})p(\boldsymbol{s}|\boldsymbol{\gamma})p(\tau),	\nonumber
	\end{align}
	where $f(\cdot)$ represents the likelihood. The algebraic form is presented in Supplementary Materials B.
Sampling from the joint posterior distribution of all parameters is obtained using Markov chain Monte Carlo (MCMC) samplers. We employ the {\tt{rnimble}} package to run the model with NIMBLE in {\tt{R}} using MCMC \citep{NIMBLE,NIMBLEr}. In the next two sections we provide examples of how our method can be applied to a range of different scenarios. All analyses have been implemented in \texttt{R} code that will be available online upon publication.
\section{Simulation study}
\label{ChapSimu}
\begin{figure}[H]
	\centering
	\includegraphics[,width=0.65\linewidth]{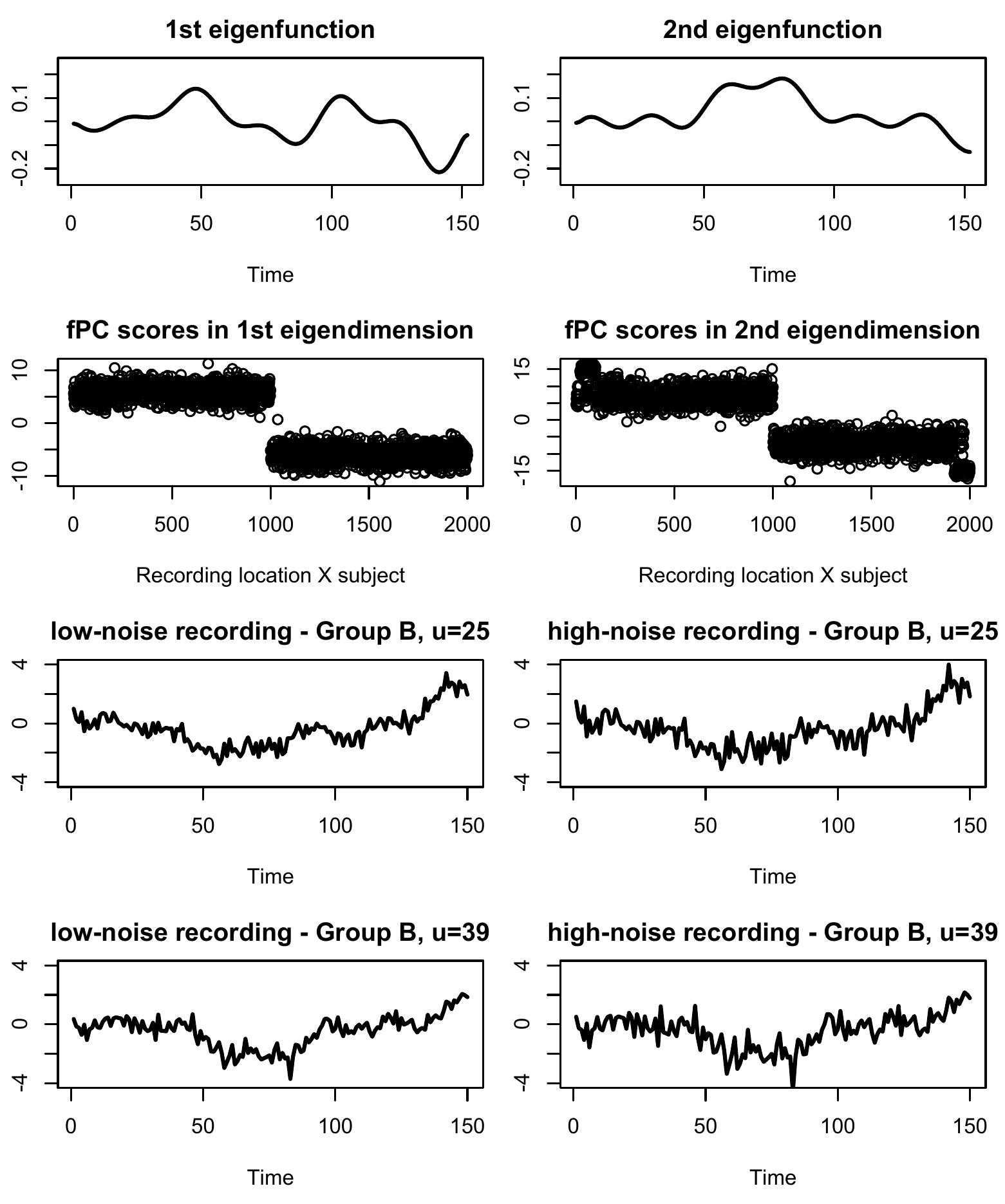}
	\caption{Simulation study: eigenfunctions  (1st row), fPC scores used for all subjects and recording locations (2nd row), and two examples of the synthetic recordings obtained with different noise levels (3rd and 4th rows).}
	\label{Fig2} 
\end{figure}

We perform a Monte Carlo simulation study to assess the performance of the proposed model in curve classification between common, group and subject-specific clusters for different levels of noise in the data.
We consider a simulation setting consisting of 40 subjects enrolled in two equally-sized study groups, denoted A and B. To generate the spatio-temporal datasets for all individuals, we employ two eigenfunctions (Figure~\ref{Fig2}, top row) developed to generate brain-like patterns (bottom row of Figure~\ref{Fig2}) including an event-evoked waveform and additional features to modulate heterogeneity in subjects and groups. For each subject $u=1,\dots,40$, we generate data in the form of $ t=1,\dots,150 $ time-points for $ n=1,\dots,50 $ recording locations. The dimensionality of the overall dataset is therefore $ 40\times 50 \times 150 $.
Group clusters are assumed to be different between groups in both eigendimensions but, within each of the two groups, we select a small subset of individuals to have outlying patterns in their recordings as generated by a different structure in their fPC scores distribution. Specifically, two subjects in group A and B ($u=1,2,39,40$) present two clusters in their fPC scores in the second eigendimension (Figure~\ref{Fig2}). Recording device-specific noise sources and artefacts are assumed to be treated in a pre-processing step and we apply a random Gaussian noise and consider models with two different noise levels (low noise: signal-to-noise ratio (SNR)=6 and  high noise: SNR=2). The bottom plots in Figure~\ref{Fig2} show an example of the curves obtained by adding either low or high random noise to the smooth data. The hyperparameter values used to run the model are listed in Supplementary Material A. The MCMC simulations were run for  $20\times 10^3$ iterations and the first $10\times 10^3$ were discarded as burn-in (This appeared to be a conservative estimate for the burn-in). The average computational time was 50 minutes (SD= 3 minutes) as measured by a computer with 8 cores, 64 GB RAM and a M1 processor. We generated 100 datasets to use in our study.

Clustering performance was assessed using two different methods: the Adjusted Rand Index (ARI) and the Variation of Information (VI) \citep{hubert1985comparing,wade2018bayesian}. The ARI is 
used in the literature to evaluate clustering performance using maximum a posteriori probabilities to quantify the similarity between the estimated partitions and the ground truth \citep{hubert1985comparing,mcdowell2018clustering}. We also employed similarity matrices to obtain a clustering point estimation via the VI approach.This method overcomes known limitations of the ARI, namely a tendency of the optimal partition to overestimate the number of clusters. In addition, we make use of credible balls to explore the uncertainty around the partition point estimate. A credible ball summarizes the posterior uncertainty around a clustering estimate and is defined as the smallest ball with posterior probability of at least 1-alpha. Further details can be found in \citet{wade2018bayesian}.
\subsection{Simulation results}
\label{Chap32}
In this section, we report the results of the simulation study for the two noise-level scenarios and eigendimensions.\vspace*{6pt}

\textit{Low noise, eigendimension 1:} there were 2 group-specific clusters in the first eigendimension. Classification using ARI produced correct classification (ARI=1) in 93 out of 100 simulations. For the other seven simulated datasets, the model returned an ARI of 0.95 and 0.91 in four and three datasets corresponding to 1 and 2 subject misclassified, respectively. Using VI, correct classification was obtained in 97 simulations while 1 subject was misclassified in the remaining 3 simulations.
Variation measured with 95\% cluster balls indicated very low cluster uncertainty: only five of the 97 simulations that correctly classified the subjects identified alternative partitions within the 95\% cluster ball
where the fPC scores of either one or two subjects in group B were allocated to subject-specific clusters.

\textit{Low noise, eigendimension 2:} there were 2 group-specific and 4 subject-specific clusters in the second eigendimension. Classification using both ARI and VI returned the correct partition in all 100 simulations. 
The 95\% cluster balls indicated very low cluster uncertainty: only one simulation identified an alternative classification where one additional subject in group B was allocated to subject-specific clusters. Subject-specific clustering of the 50 recording locations returned low classification error with two third of the simulated datasets returning zero or a single classification error and two or three errors in one third of the datasets.

\textit{High noise, dimension 1:} under high noise (STN=2), ARI returned exact classification (ARI=1) in 62 out of 100 simulations, misclassified one subject in 23 simulations (ARI= 0.95), two subjects in 14 simulations (ARI= 0.90) and three subjects in 1 simulation (ARI=0.87). Using VI, all individuals were correctly classified in 77 simulations; one individual was misclassified in 15 simulations; and two individuals were misclassified in 8 simulations. The three misclassified subjects were among those with individual-specific clusters in the second eigendimension. 
Five simulations with misclassified individuals included the true partition within their 95\% cluster ball.
The 95\% cluster balls indicated very low cluster uncertainty: only  5 of the 77 simulations that correctly classified the subjects identified alternative partitions within the 95\% cluster ball where either one or two subjects in group B were allocated to subject-specific clusters.

\textit{High noise, dimension 2:} Classification using both ARI and VI returned correct classification in all 100 simulations. The 95\% cluster balls indicated again very low cluster uncertainty: only two simulations identified alternative classifications where one additional subject in group B was allocated to subject-specific clusters. Subject-specific clustering of the 50 recording locations returned again low classification error with 60\% and 40\% of the simulated datasets returning zero to a single classification error and two to three errors, respectively.

Overall, the model showed very good to optimal performance in retrieving the correct partitions in both eigendimensions and different noise levels, particularly when the VI was used. When the correct partition was identified, there was always low cluster uncertainty around it, with only one or two additional subjects allocated to subject-specific clusters within the 95\% cluster ball. A higher noise level led only to a moderate increase in the misclassification for both subject-level and recording-level partitions. In the next section, we apply our model on a different challenging scenario using a real neuroscientific dataset which featured both a smaller sample size and a potentially higher inter-individual variability in the recordings.  
\section{Case study: Event-related potentials}
\label{ChapERPs}
We consider data from an EEG study on brain activations following object recognition tasks \citep{zhang1995event}. Event-Related Potentials (ERPs) continue to be a popular electrophisiological tool to investigate memory functions under both physiological and pathological conditions (see, for example, \citet{yan2023effect,saltzmann2022neural,stevens4208065relational}). Small bio-electrical signals are generated by the brain in response to specific events or stimuli. These signals are time locked to cognitive events thus providing a non-invasive approach to study
psychophysiological correlates of mental processes \citep{sur2009event}. Relevant demographics and clinical features were controlled for at the design stage. Subjects were shown two separate images taken from the 1980 Snodgrass and Vanderwart picture \citep{snodgrass1980standardized}. The second stimulus was either a different image (unmatched) or the same image (matched). The data were recorded using a cap with 64 electrodes placed on the subject's scalp and the brain activity at each recording electrode was sampled at 256 Hz for 1 second. Further details on the enrolled cohort and the recording setting can be found in \citet{zhang1995event}. 

We focus our analysis on the first 300 milliseconds (77 time points) recorded in the occipital region of the brain where the visual cortex is located \citep{crossman2018neuroanatomy}. In particular, we explore whether our approach offers additional insights into the spatio-temporal differences and the inter-subject variability in the ERPs recorded from this area. 

\begin{figure}
	\centering
	\includegraphics[width=0.65\linewidth]{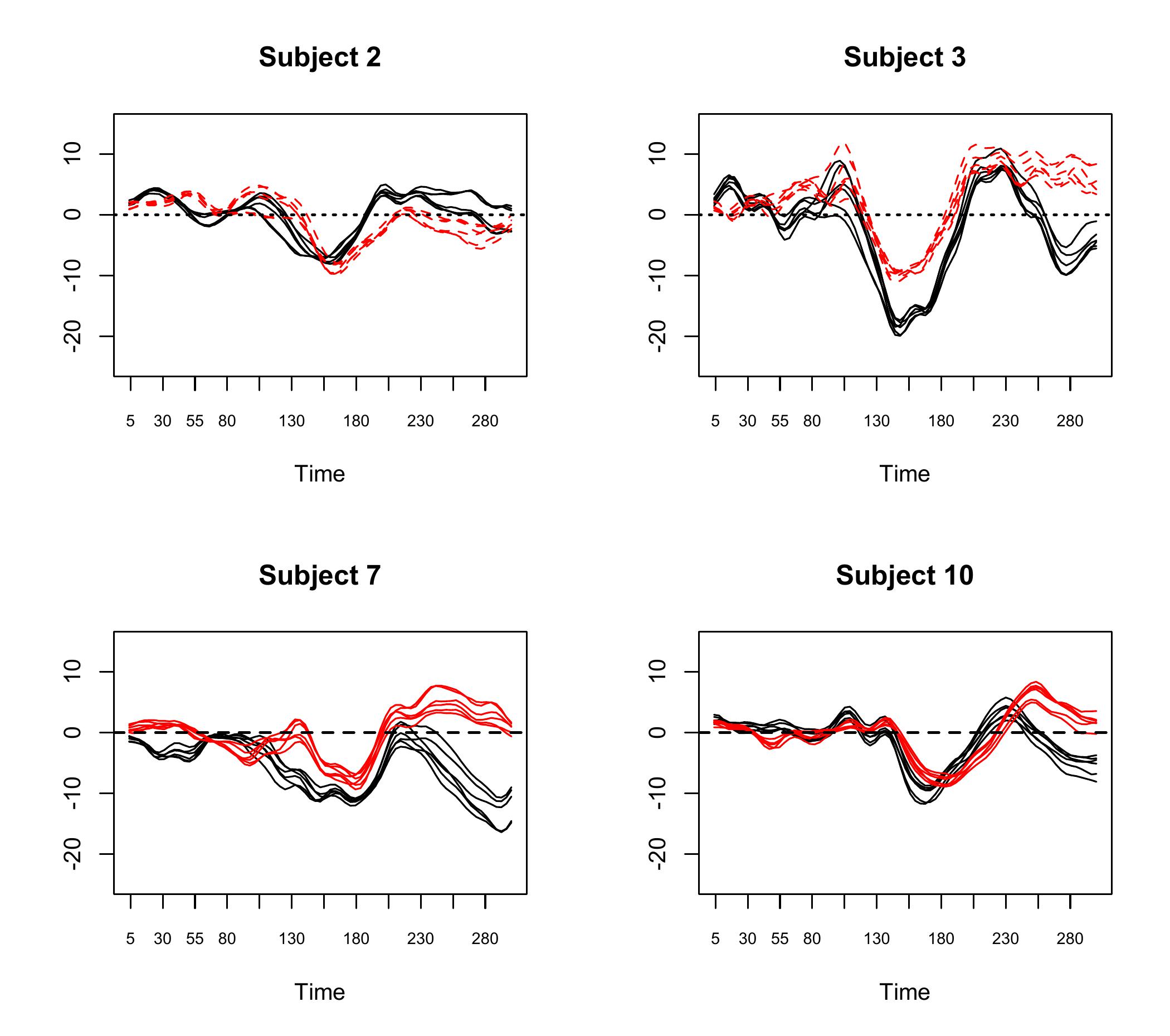}
	\caption{ERPs study: an example of recordings from four subjects included in the analyses. ERPs recorded during matched stimuli are in black (solid), those recorded during unmatched stimuli are in red (dashed).}
	\label{Fig3} 
\end{figure}

For each subject, multiple repetitions of the tasks are averaged to amplify the underlying evoked response. The resulting potentials are visually inspected to ascertain the presence of the evoked signals and eliminate recordings affected by artifacts. Three components are expected at approximately 110, 175 and 247 milliseconds (named C110, C175 and C247) \citep{zhang1995event}. Following standard terminology, an epoch identifies a time-window extracted from the continuous EEG signal of an individual under either condition while negative and positive ERP peaks identify convex and concave trajectories, respectively. The final dataset includes 20 epochs, 60 recordings per stimulus from 10 subjects (i.e. six trajectories per subject). Smoothed data of 4 individuals are shown in Figure \ref{Fig3}, where ERPs following matched stimuli and unmatched stimuli are shown in black (solid) and red (dashed), respectively. The complete dataset is shown in Supplementary Materials D. 
A final ($120\times 77$) dataset is input to fPCA for curve smoothing and dimension reduction using the \texttt{pca.fd} function from the \texttt{fda} package in \texttt{R} \citep{fdapackage}. Due to the multilevel structure of the data, trail eigendimensions explaining little variation are most likely to capture subject-specific rather than group trends; therefore, we retained the first three eigendimensions explaining almost $ 80\% $ of the total variability while accounting for more than $ 15\% $ each. The smoothed recordings and the first two retained eigendimensions are shown in Figure~\ref{Fig4} while the third eigendimension is shown in Figure 2 of Supplementary Materials D. We used the model settings described in Section~\ref{Chap2} while in Supplementary Materials C we present the result of a sensitivity analysis where the impact of choosing different hyperparameter values is explored. Convergence was assessed using trace plots and Brooks-Gelman-Rubin diagnostics and chain length was monitored using the effective sample size (see Supplementary Materials E for further details). The computational time to run $2\times 10^5$ iterations (discarding the first $10^5$ as burn-in) using NIMBLE on a Quad-core Intel CPU running at 2 GHz with 16 GB RAM was 90 minutes.
\subsection{ERPs analysis results}
\label{Chap41}
\begin{figure}
	\centering
	\includegraphics[width=0.9\linewidth]{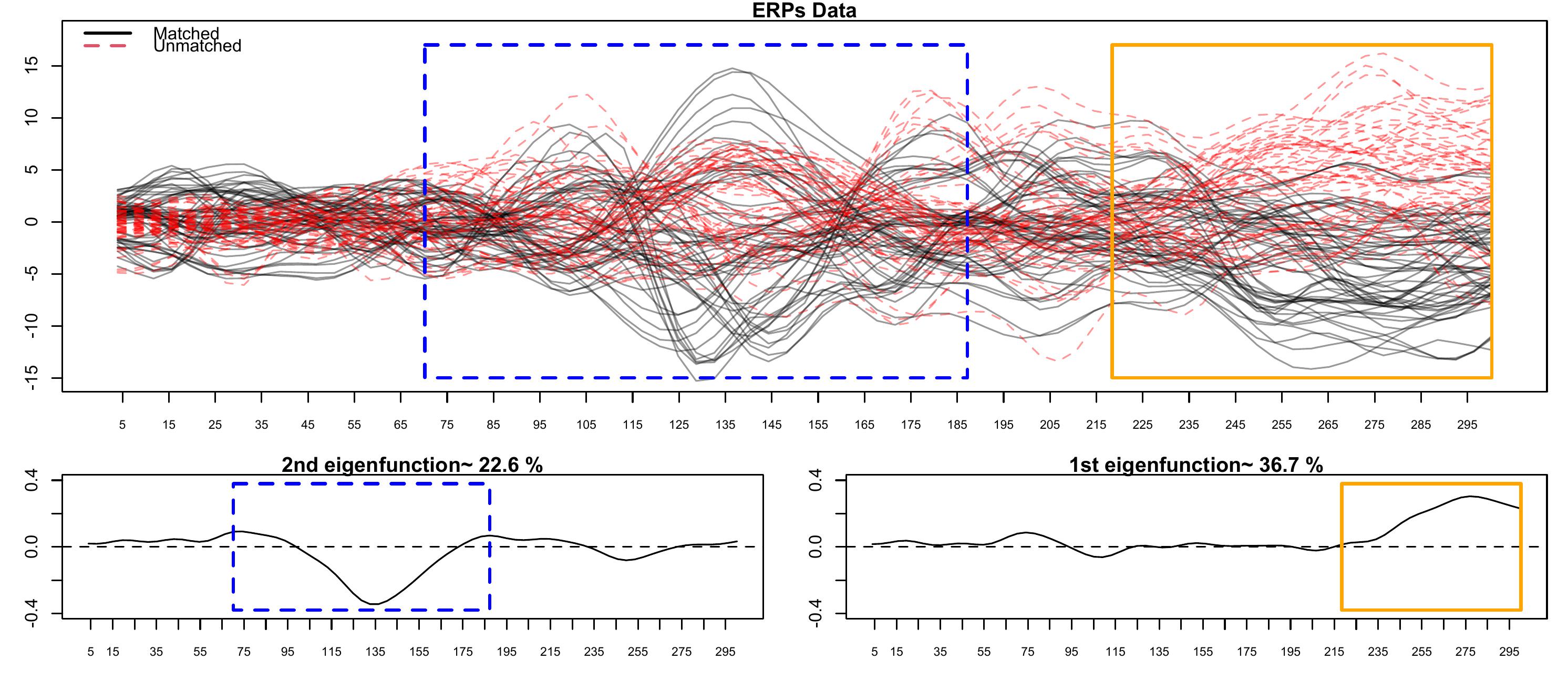}
	\caption{ERPs study: matched (solid) and unmatched (dashed) trajectories. The boxes highlight areas of higher variability described by the first (solid) and second (dashed) eigendimensions.}
	\label{Fig4} 
\end{figure}

\begin{figure}
	\centering
	\includegraphics[width=0.8\linewidth]{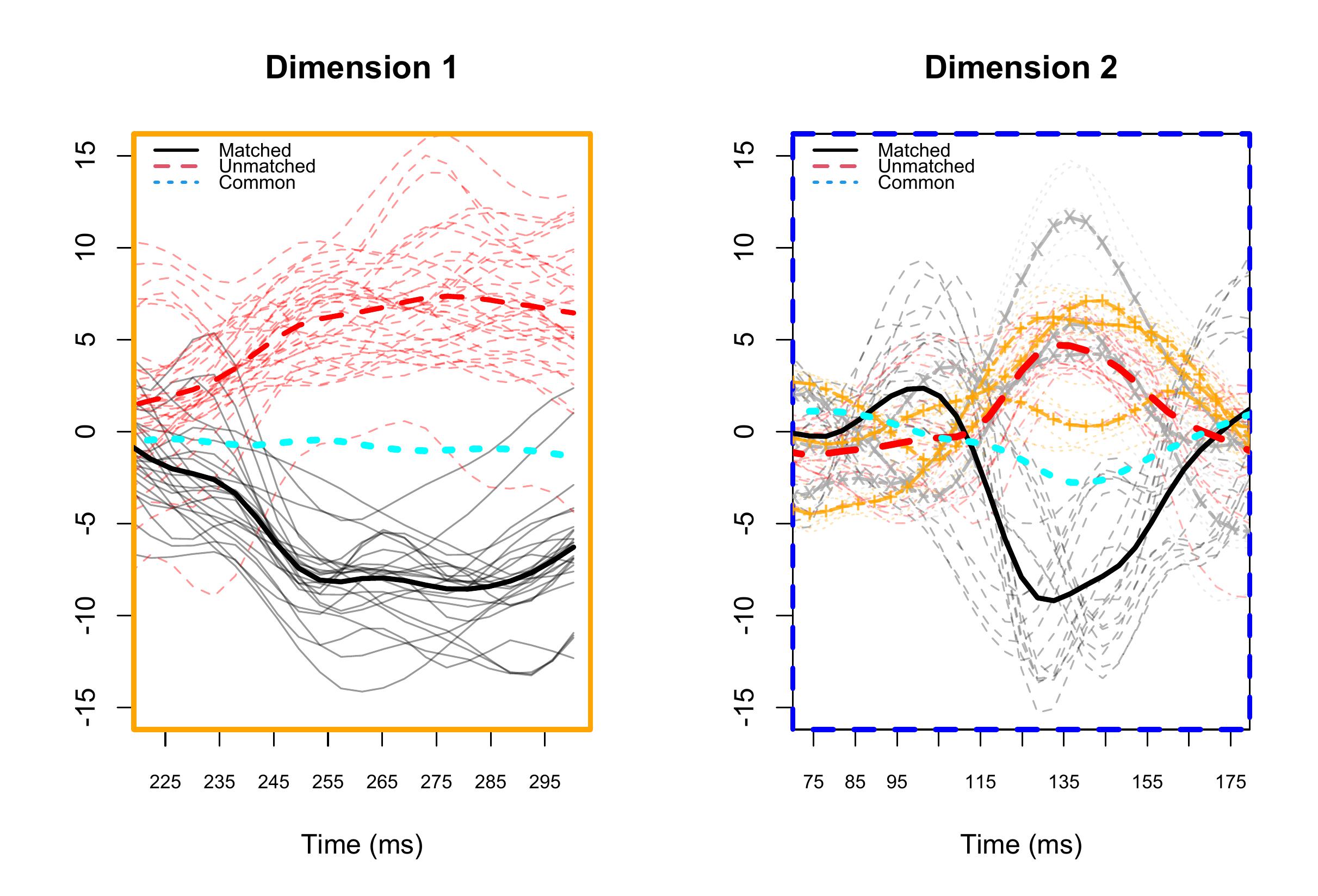}
	\caption{ERPs study: local group differences between matched (solid) and unmatched (dashed) recordings identified by the model in the first two eigendimensions. In bold the common and group cluster means are superimposed on the relative recordings. The symbols x (grey) and '$+$'(orange) identify individual-specific cluster means. }
	\label{Fig5} 
\end{figure}

The results of the ERPs study are presented separately for each of the three eigendimensions analysed to gain insights into local features of the recordings. For each dimension, we explore how subjects' recordings were allocated between common, group, and subject-specific clusters. Table \ref{Table1} reports the partitions obtained in each eigendimension and their relative 95\% cluster balls. The three point estimates take most of the variation within the relative cluster balls with only minor or negligible deviations in the first and the other two eigendimensions, respectively.\vspace*{6pt}\\
\begin{table}
	\caption{ERPs study: the 3 panels show the partitions obtained within each of the 3 eigendimensions studied (C=common, Gm=matched group, Gu=unmatched group, S=subject-specific). Each panel reports the point estimate and the relative three bounds of the 95\% credible ball (upper, lower, horizontal; see \citet{wade2018bayesian} for further details). Within each bound, the differences with the point estimate are highlighted in bold. At the bottom of each panel it is shown the frequency of the different partitions found within the credible ball. \label{Table1}}
	%	\small\sf
	\sf
	\begin{center}
		\resizebox{11cm}{!}{%
			\begin{tabular}{ll|ll|ll|ll}
				\hline
				\multicolumn{8}{c}{\textbf{1st eigendimension}}\vspace{5pt}\\ 
				\multicolumn{2}{l|}{Point estimate} & \multicolumn{2}{l|}{95\% CB v. upperb.} & \multicolumn{2}{l|}{95\% CB v. lowerb.} & \multicolumn{2}{l}{95\% CB horizb.} \vspace{4pt} \\ \hline
				C: 	& 1m, 2m, 3m,          &           &            & C:          & 1m, 3m, 4m,        & C:           &  1m, 3m, 4m,        \\
				&4m, 9m, 10m,          &           &           &           & 9m, 10m, 1u,         &           &9m, 10m, 1u,        \\
				&1u, 2u, 5u,          &           &           &           &2u, 5u, 6u  &           &2u, 5u, 6u  \\  
				&6u           &           &           &           &          &           & \vspace{5pt} \\  
				Gm:	&5m, 6m, 7m,    &           &$\boldsymbol{\equiv}\text{Point estim.}\quad\,\,\, $          & Gm:          &  5m, 6m, 7m, $\quad$          &Gm:           & 5m, 6m, 7m,  \\ 
				&8m     &           &        &           &8m  &  &8m \vspace{5pt} \\ 
				Gu:	&3u, 4u, 7u,      &           &           &Gu:           & 3u, 4u, 7u,          &Gu:           & 3u, 4u, 7u, \\
				&8u, 9u, 10u           &           &           &           &8u, 9u, 10u&           &8u, 9u, 10u \vspace{5pt} \\ 
				S:	&$ \emptyset $           &           &           & S:          &  \textbf{2m}         & S:          & \textbf{2m}  \vspace{5pt} \\ \hline
				& Frequency:        &          &     $100\%$     &           & $ 6\% $           &           &  $ 6\% $   \vspace{0.7cm}  
			\end{tabular}
		}
		\resizebox{11cm}{!}{%
			\begin{tabular}{ll|ll|ll|ll}
				\hline
				\multicolumn{8}{c}{\textbf{2nd eigendimension}}\vspace{5pt}\\ 
				\multicolumn{2}{l|}{Point estimate} & \multicolumn{2}{l|}{95\% CB v. upperb.} & \multicolumn{2}{l|}{95\% CB v. lowerb.} & \multicolumn{2}{l}{95\% CB horizb.} \vspace{4pt}   \\  \hline 
				C: 	& 2m, 6m, 7m,          & C:          & 2m, 7m, 9m,         & C:          &7m, 9m, 5u,          & C:           &7m, 9m, 3u,          \\
				&9m, 3u, 5u,           &           &3u, 5u, 6u,          &           &6u,8u          &           &5u, 6u, 8u       \\  
				& 6u, 8u         &           & 8u, \textbf{9u}         &           &         &           &        \vspace{5pt}    \\  
				Gm:	&3m, 5m, 8m     &  Gm:         & 3m, 5m, \textbf{6m},          & Gm:          &$\emptyset$            &Gm:           &3m, 5m, \textbf{6m}\\ 
				&    &          & 8m          &         &          &          &  \vspace{5pt} \\ 
				Gu:	&2u, 4u, 7u       & Gu:          & 2u, 4u, 7u,         &Gu:           & 2u, 4u, 7u          &Gu:           &2u, 4u, 7u, \\ 
				&    &           &\textbf{10u}          &          &           &           &\textbf{9u}  \vspace{5pt} \\ 
				S:	&1m, 4m, 10m,           & S:          &1m, 4m, 10m,           & S:          &1m, \textbf{2m}, \textbf{3m},         & S:          &1m, \textbf{2m}, 4m,	 \\
				&1u, 9u, 10u           &           & 1u          &           &4m, \textbf{5m}, \textbf{6m},         &           &\textbf{8m}, 1u, 10u	\\
				&          &           &           &           & \textbf{8m}, 10m, 1u,         &           &	 \\ 
				&          &           &           &           & \textbf{3u}, 9u, 10u         &           &		 
				\vspace{5pt} \\ \hline
				& Frequency:        &          & $< 1\% $          &           & $ < 1\% $           &           &  $ 1\% $   \vspace{0.7cm}  
			\end{tabular}
		}
		\resizebox{11cm}{!}{%
			\begin{tabular}{ll|ll|ll|ll}
				\hline
				\multicolumn{8}{c}{\textbf{3rd eigendimension}}\vspace{5pt}\\ 
				\multicolumn{2}{l|}{Point estimate} & \multicolumn{2}{l|}{95\% CB v. upperb.} & \multicolumn{2}{l|}{95\% CB v. lowerb.} & \multicolumn{2}{l}{95\% CB horizb.} \vspace{4pt} \\ \hline
				C: 	& 2m, 8m, 10m,          & C:          &\textbf{1m}, 2m, \textbf{3m},          & C:          &10m, 2u, 7u,         & C:           &\textbf{1m}, 8m, 10m,          \\
				&2u, 7u, 8u,         &           &8m, 10m, 8u           &           &9u            &           & 2u, 7u, 9u,      \\  
				&9u          &           &          &           &           &           & \textbf{10u}        \vspace{5pt}    \\  
				Gm:	&1m, 4m, 5m,     &  Gm:         & 4m, 5m, 6m,          & Gm:          &5m, 7m, 9m           &Gm:           &5m, 6m, 7m,   \\ 
				&6m, 7m, 9m    &           &7m, 9m           &           &           &          &9m     \vspace{5pt} \\ 
				Gu:	&1u, 4u, 5u,       & Gu:          &1u, \textbf{2u}, 4u,           &Gu:           &5u, 6u, 10u           &Gu:           &1u, 4u, 5u, \\ 
				&6u, 10u      &           &5u, 6u, \textbf{7u},          &           &           &           & 6u   \\ 
				&     &           &9u, \textbf{10u}           &           &           &           &  \vspace{5pt} \\ 
				S:	&3m, 3u          & S:          &3u           & S:          &\textbf{1m}, \textbf{2m}, 3m,         & S:          &\textbf{2m}, 3m, \textbf{4m},\\
				&          &           &          &           &\textbf{4m}, \textbf{6m}, \textbf{8m},         &           &3u, \textbf{8u} \\
				&          &           &          &           &\textbf{1u}, 3u, \textbf{4u},       &           & 	\\ 
				&          &           &          &           &\textbf{8u}        &           & 		 
				\vspace{5pt} \\ \hline
				& Frequency:        &          & $< 1\% $          &           & $< 1\% $           &           &  $ < 1\% $     
			\end{tabular}
		}
	\end{center}
\end{table}
\textit{Dimension 1:} This eigendimension is mostly related to variability of approximately the last 70 ms of the recording (Figure~\ref{Fig4}, bottom-right panel). Our model allocated 50\% of the subjects' recordings to group clusters (10 epochs, 4 in the matched and 6 in the unmatched condition) with a positive peak in the matched group and a negative peak in the unmatched group (Figure \ref{Fig5}, left panel). The recordings of the remaining 50\% of individuals were included in the common cluster. No subject-specific pattern was detected. This result was particularly robust to changes in the hyperparameter values (see sensitivity analysis in Supplementary Materials C).

\textit{Dimension 2:} The second eigendimension represents local variability between 75-175 ms (Figure ~\ref{Fig4}, bottom-left panel). The model allocated 30\% of the subjects' recordings to group clusters (6 epochs, 3 for each condition) with a positive peak in the matched group and a negative in the unmatched group (Figure \ref{Fig5}, right panel). The recordings of 8 subjects were allocated to the common cluster (4 for each condition) while the remaining recordings of 6 individuals were allocated to subject-specific clusters (3 for each condition); of these, three subjects in the matched condition presented an opposite peak to that of the respective group cluster while three subjects in the unmatched condition showed the same pattern of their group but with extremely pronounced peaks.

\textit{Dimension 3:} The third eigendimension includes local variability at 90-130 ms and 180-250 ms (Figure 2 in Supplementary Materials D). The recordings of 55\% subjects were allocated to group clusters (11 epochs, 6 in the matched condition and 5 in the unmatched condition) with a positive peak in the matched group and a slightly less pronounced positive peak in the unmatched group. The recordings of the other 35\% and 10\% of subjects were allocated to common and subject-specific clusters, respectively (Figure 3 in Supplementary Materials D). The recordings of the two individuals allocated to individual-specific clusters showed an inverted peak in the recordings compared to that of the respective group cluster.

Overall, dimensions 1 and 3 presented the largest group clusters, including the recordings of 50\% and 55\% of subjects respectively, and the fewest individual-specific clusters (only 2 subjects in dimension 3). Conversely, dimension 2 showed the smallest group clusters (including only 6 subjects) and the largest number of individual-specific clusters (6 subjects). The latter result was also confirmed in the sensitivity analysis (Supplementary Materials C). These results are in agreement with the original study of \citet{zhang1995event} but provide further understanding on the heterogeneity of ERP responses in the occipital lobe, following matched vs unmatched visual stimuli. The group patterns identified in the first dimension capture the difference in the latency of the component at 247ms (C247) between matched and unmatched trials as found by  \citet{zhang1995event} and in a similar experiment also by \citet{begleiter1993neurophysiologic} This pattern explained approximately $40\%$ of the total variability in our sample.

Results of our analysis suggested that stimulus-specific brain responses in the occipital lobe became highly synchronised between 230ms and 300ms. However, only half of the recordings were allocated to group-specific clusters while the other 50\% were pooled together in the common cluster, suggesting that the effect of non-matching stimuli on the latency of the C247 component in the occipital lobe might be a trait not consistent in the population.

The group pattern identified in the second dimension corresponds to a longer latency of the C110 component during non-matching stimuli. This result was not found in \citet{zhang1995event} but it is in agreement with the study of \citet{begleiter1995event} In our analysis, the highest individual heterogeneity in the trajectories was found associated to this component, with 30\% of the subjects' recordings being classified in  subject-specific clusters (which might explain why the group effect was not originally identified by \citet{zhang1995event}). Interestingly, subject-specific clusters in the matched group included markedly inverted peaks while those in the unmatched group had much less extreme patterns, suggesting large heterogeneity around the C110 component might be mostly elicited by the matched condition.

In the third eigendimension, half of the recordings were again assigned to group clusters although the clusters means have  different magnitude but the same sign. Notably, one subject's recordings were allocated to subject-specific clusters under both conditions (Subject 3, see Figure 1 in Supplementary Materials D). As the variability explained by this eigendimension is only 17\% of the total variation, most of this variation could have been generated by a single individual, as Figure 3 in Supplementary Materials D also suggests. Considering that only the initial eigendimensions capture general patterns in the data, these results suggest limiting the interpretation of our model output to the first two eigendimensions.

In summary, our results indicate that group differences in occipital ERPs elicited by the specific visual tasks analysed arise at a local level (dimension 1 and 2) rather than over the whole recording and could be detected only on subsets of the individuals analysed, the largest being 10 individuals (50\%) in the first dimension. Moreover, we found that heterogeneity in subjects' brain responses change over the length of the EEG recordings. It is highest and presents stimulus-specific differences between 75ms and 175ms (dimension 2), while it is lowest after 230ms (dimension 1). 
The identification of these complex spatio-temporal patterns highlights the insights that can be gained by exploring brain recordings in group studies using our proposed method.
\section{Discussion}
We introduce a novel approach based on a Bayesian functional Principal Components Analysis model and a modified Dirichlet Process mixture prior to explore the spatio-temporal features of neuroscientific recordings and investigate inter-subject heterogeneity in the analysis of groups.

The proposed method allows for an in-depth exploration of local patterns arising from complex spatio-temporal data recorded from experimentally designed groups (e.g. patients vs controls, pre vs post task/treatment). By using the eigenfunctions as smooth constituent blocks of the recorded signal, we focused the analyses on only the most important temporal features in the data. The data-driven trade-off between subject-level and recording-level partitions of the fPC scores allows for a flexible exploration of both group-specific behaviours and inter-individual variability. This returns an in-depth analysis of the whole multilevel spatio-temporal dataset where a subject might share some of the temporal features of their recordings either with individuals from all groups or from their own group only or with no one else.

Simulation results showed that the method proposed returns correct classification, with both low and high noise levels in the data with the latter introducing some minor degree of misclassification. As expected, using the Variation of Information proposed by \citet{wade2018bayesian} produced more reliable classification under both STN ratios, compared to the Adjusted Rand Index. Interestingly, the misclassification observed affected the first eigendimension only. We believe this could be the result of using a global smoothing approach on non-homogeneous dependent curves affected by high noise. The  consequence for classification observed was a tendency to assign one or two individuals to subject-specific clusters beyond the eigendimension where this allocation was correct. Therefore, smoothing at group or subject level could be employed to account for differences in STN ratio among these. Local smoothing could also be considered to account for heteroschedasticity in the recordings. Several local smoothers are available (e.g. local Cross Validation, local Generalised Cross Validation) \citep{loader2006local}. Although these methods are considered challenging especially when high noise and complex structures characterise the data, they are still an active area of research \citep{grzesik2017local}.

The results of the case study on ERPs data not only were in agreement with the relevant literature \citep{zhang1995event,begleiter1995event,begleiter1993neurophysiologic}, but also offered additional insights on the underlying complex spatio-temporal process. In particular, we were able to find group differences between matched/unmatched stimuli and link them to specific temporal features. In addition, by looking at the composition of the different clusters we uncovered how individual heterogeneity varied across the main temporal modes of variation in the data.The identified heterogeneity in subjects' responses was in line with works on subjects' heterogeneity in brain responses to recognition tasks \citep{Dimsdalezucker2022108287,Turano20161,LaszloO201542}. It is worth noting that the repeated recordings used were averaged in the pre-processing step to extract reliable ERPs. This has been the standard approach in many past and recent ERPs studies and we follow it here for comparability with the relevant literature \citep{zhang1995event,begleiter1993neurophysiologic,begleiter1995event}. Nonetheless, the approach assumes the variability in the repeated tasks to be negligible while alternative approaches are available (see \citet{mouraux2008across} for a discussion of different methods). In this respect, our new modelling approach could be used in the exploration of individual variation due to repeated measurements, where interest lies in identifying local features of individual-specific repeated trajectories that deviate from those of the relative group. Applying our approach would permit a local rather than global exploration of individual variations due to the repeated design, whilst ensuring that comparisons between groups are not affected by outlying features.

From a practical point of view, our method requires a series of pre and post analyses checks (tuning of hyperparameters values, MCMC checks, sensitivity analysis) that go beyond those required for standard ANOVAs and Linear Mixed Models. Nevertheless, as recording of large and complex datasets is becoming the standard practice, researchers can take a deeper look into the mechanisms governing many biological phenomena as long as they adopt methods that can capture the complexity of such mechanisms. In this respect, our method offers a much more intricate but complete picture of the underlying brain activity compared to the results of the original studies.

There are many interesting research avenues that can be taken to expand this methodology further. For example,known covariates can be accounted for by considering a linear modelling of the fPC scores cluster means for the group clusters. Further, more intricate solutions are envisaged for the case with several covariates, for example, by considering using a mixture of regressions and implementing a variable selection approach \citep{cozzini2014bayesian}.
Another important challenge in neuroscience is performing group comparison when data is high dimensional and/or on multiple domains (e.g. both time and frequency domain data). In this scenario, incorporating recent developed fPCA methods (e.g., fPCA for images \citep{shi2022two} or for the analysis of data observed on different domains \citep{happ2018multivariate}) in our model is a promising approach to tackle this challenge and is an area of future research.

A further research direction concerns groups. In our multilevel model we considered two known groups such as patients and controls or treatment A and B. However, there are many cases where groups are unknown and interest may lie in finding them. For example in the identification of different courses in diseases with heterogeneous manifestations as well as in the identification of prognostic patterns in multimorbidity studies. In addition, considering the whole brain, as opposed to targeted brain areas, may return insightful clustering at both individual and group-specific levels. To tackle these challenges it is possible to extend our model via a variety of multilevel Dirichlet Process structures such as the Nested DP \citep{rodriguez2008nested}, the Hierarchical DP \citep{teh2006w} or the Multilevel clustering Hierarchical DP \citep{wulsin2012hierarchical}.

Another timely research direction concerns the applicability of the proposed method to the analysis of large to very large datasets. Given the current trend of technological advancements, the fast accumulation of large quantities of information and their easy access, it is essential to develop scalable methodologies that can compete computationally with machine learning approaches while permitting a more complex analysis of patterns and uncertainty in the data. In this regard, divide-and-conquer and subsampling methods represent a promising research areas for the study of large datasets (see for example \citet{king2022large,su2020divide,song2020distributed,quiroz2018speeding}). In our future research we aim to explore how these methods could be adapted to implement our modelling strategy on large neuroscientific datasets.

\vspace*{1pc}

\noindent {\bf{Acknowledgment}}
\noindent {In order to meet institutional and research funder open access requirements, any accepted manuscript arising shall be open access under a Creative Commons Attribution (CC BY) reuse licence with zero embargo.}\\
\noindent {\bf{Research Ethics}}
\noindent {Not applicable.}\\
\noindent {\bf{Informed Consent}}
\noindent {Not applicable.}\\
\noindent {\bf{Authors Contributions}}
\noindent {All authors have accepted responsibility for the entire content of this manuscript and approved its submission.}\\
\noindent {\bf{Conflict of Interest}}
\noindent {The authors state no conflict of interest.}\\
\noindent {\bf{Research Funding}}
\noindent {This study was supported by an Early Career Fellowship of the London Mathematical Society (LMS) [ECF-1920-40].}\\
\noindent {\bf{Data Availability Statement}}
\noindent {The reproducible R code for simulations in this paper will be provided on the GitHub repository\\ {\tt{https://github.com/NicMargaritella/BfPCA\_MPP}} upon publication. Event Related Potentials data used in this paper to support our findings is available from the UC Irvine Machine Learning Repository website\\ (\tt{DOI: https://doi.org/10.24432/C5TS3D}).}\\
\noindent {\bf{Supplementary Materials}}
\noindent {This article contains supplementary material.}
\bibliographystyle{unsrtnat}
\bibliography{References.bib}

\begin{thebibliography}{61}
\providecommand{\natexlab}[1]{#1}
\providecommand{\url}[1]{\texttt{#1}}
\expandafter\ifx\csname urlstyle\endcsname\relax
  \providecommand{\doi}[1]{doi: #1}\else
  \providecommand{\doi}{doi: \begingroup \urlstyle{rm}\Url}\fi

\bibitem[Alday et~al.(2017)Alday, Schlesewsky, and
  Bornkessel-Schlesewsky]{alday2017electrophysiology}
Phillip~M Alday, Matthias Schlesewsky, and Ina Bornkessel-Schlesewsky.
\newblock Electrophysiology reveals the neural dynamics of naturalistic
  auditory language processing: event-related potentials reflect continuous
  model updates.
\newblock \emph{eNeuro}, 4\penalty0 (6), 2017.

\bibitem[Yu et~al.(2022)Yu, Guindani, Grieco, Chen, Holmes, and
  Xu]{yu2022beyond}
Zhaoxia Yu, Michele Guindani, Steven~F Grieco, Lujia Chen, Todd~C Holmes, and
  Xiangmin Xu.
\newblock Beyond t test and anova: applications of mixed-effects models for
  more rigorous statistical analysis in neuroscience research.
\newblock \emph{Neuron}, 110\penalty0 (1):\penalty0 21--35, 2022.

\bibitem[Hasson and Honey(2012)]{hasson2012future}
Uri Hasson and Christopher~J Honey.
\newblock Future trends in neuroimaging: Neural processes as expressed within
  real-life contexts.
\newblock \emph{NeuroImage}, 62\penalty0 (2):\penalty0 1272--1278, 2012.

\bibitem[Ramsay and Silverman(2005)]{ramsay2005functional}
James Ramsay and Bernard~Walter Silverman.
\newblock \emph{Functional {D}ata {A}nalysis}.
\newblock Springer Series in Statistics, 2005.

\bibitem[Gertheiss et~al.(2023)Gertheiss, R{\"u}gamer, Liew, and
  Greven]{gertheiss2023functional}
Jan Gertheiss, David R{\"u}gamer, Bernard~XW Liew, and Sonja Greven.
\newblock Functional data analysis: An introduction and recent developments.
\newblock \emph{arXiv preprint arXiv:2312.05523}, 2023.

\bibitem[Viviani et~al.(2005)Viviani, Gr{\"o}n, and
  Spitzer]{viviani2005functional}
Roberto Viviani, Georg Gr{\"o}n, and Manfred Spitzer.
\newblock Functional principal component analysis of f{MRI} data.
\newblock \emph{Human {B}rain {M}apping}, 24\penalty0 (2):\penalty0 109--129,
  2005.

\bibitem[Hasenstab et~al.(2017)Hasenstab, Scheffler, Telesca, Sugar, Jeste,
  DiStefano, and {\c{S}}ent{\"u}rk]{hasenstab2017multi}
Kyle Hasenstab, Aaron Scheffler, Donatello Telesca, Catherine~A Sugar, Shafali
  Jeste, Charlotte DiStefano, and Damla {\c{S}}ent{\"u}rk.
\newblock A multi-dimensional functional principal components analysis of {EEG}
  data.
\newblock \emph{Biometrics}, 73\penalty0 (3):\penalty0 999--1009, 2017.

\bibitem[Zhang and Parnell(2023)]{zhang2023review}
Mimi Zhang and Andrew Parnell.
\newblock Review of clustering methods for functional data.
\newblock \emph{ACM Transactions on Knowledge Discovery from Data}, 17\penalty0
  (7):\penalty0 1--34, 2023.

\bibitem[Angelini et~al.(2012)Angelini, De~Canditiis, and
  Pensky]{angelini2012clustering}
Claudia Angelini, Daniela De~Canditiis, and Marianna Pensky.
\newblock Clustering time-course microarray data using functional {B}ayesian
  infinite mixture model.
\newblock \emph{Journal of {A}pplied {S}tatistics}, 39\penalty0 (1):\penalty0
  129--149, 2012.

\bibitem[Ray and Mallick(2006)]{ray2006functional}
Shubhankar Ray and Bani Mallick.
\newblock Functional clustering by {B}ayesian wavelet methods.
\newblock \emph{Journal of the {R}oyal {S}tatistical {S}ociety: {S}eries B
  ({S}tatistical {M}ethodology)}, 68\penalty0 (2):\penalty0 305--332, 2006.

\bibitem[Zhou and Wakefield(2006)]{zhou2006bayesian}
Chuan Zhou and Jon Wakefield.
\newblock A {B}ayesian mixture model for partitioning gene expression data.
\newblock \emph{Biometrics}, 62\penalty0 (2):\penalty0 515--525, 2006.

\bibitem[James and Sugar(2003)]{james2003clustering}
Gareth~M James and Catherine~A Sugar.
\newblock Clustering for sparsely sampled functional data.
\newblock \emph{Journal of the {A}merican {S}tatistical {A}ssociation},
  98\penalty0 (462):\penalty0 397--408, 2003.

\bibitem[Gonzalez-Castillo and Bandettini(2018)]{gonzalez2018task}
Javier Gonzalez-Castillo and Peter~A Bandettini.
\newblock Task-based dynamic functional connectivity: Recent findings and open
  questions.
\newblock \emph{Neuroimage}, 180:\penalty0 526--533, 2018.

\bibitem[Warnick et~al.(2018)Warnick, Guindani, Erhardt, Allen, Calhoun, and
  Vannucci]{warnick2018bayesian}
Ryan Warnick, Michele Guindani, Erik Erhardt, Elena Allen, Vince Calhoun, and
  Marina Vannucci.
\newblock A {B}ayesian approach for estimating dynamic functional network
  connectivity in f{MRI} data.
\newblock \emph{Journal of the American Statistical Association}, 113\penalty0
  (521):\penalty0 134--151, 2018.

\bibitem[Hutchison et~al.(2013)Hutchison, Womelsdorf, Allen, Bandettini,
  Calhoun, Corbetta, Della~Penna, Duyn, Glover, Gonzalez-Castillo,
  et~al.]{hutchison2013dynamic}
R~Matthew Hutchison, Thilo Womelsdorf, Elena~A Allen, Peter~A Bandettini,
  Vince~D Calhoun, Maurizio Corbetta, Stefania Della~Penna, Jeff~H Duyn, Gary~H
  Glover, Javier Gonzalez-Castillo, et~al.
\newblock Dynamic functional connectivity: promise, issues, and
  interpretations.
\newblock \emph{Neuroimage}, 80:\penalty0 360--378, 2013.

\bibitem[Seghier and Price(2018)]{seghier2018interpreting}
Mohamed~L Seghier and Cathy~J Price.
\newblock Interpreting and utilising intersubject variability in brain
  function.
\newblock \emph{Trends in Cognitive Sciences}, 22\penalty0 (6):\penalty0
  517--530, 2018.

\bibitem[Li et~al.(2017)Li, Yin, Zhu, Ren, Yu, Wang, Zheng, Niu, Huang, and
  Li]{li2017linking}
Rui Li, Shufei Yin, Xinyi Zhu, Weicong Ren, Jing Yu, Pengyun Wang, Zhiwei
  Zheng, Ya-Nan Niu, Xin Huang, and Juan Li.
\newblock Linking inter-individual variability in functional brain connectivity
  to cognitive ability in elderly individuals.
\newblock \emph{Frontiers in Aging Neuroscience}, 9:\penalty0 385, 2017.

\bibitem[Zhang et~al.(2016)Zhang, Guindani, Versace, Engelmann, Vannucci,
  et~al.]{zhang2016spatiotemporal}
Linlin Zhang, Michele Guindani, Francesco Versace, Jeffrey~M Engelmann, Marina
  Vannucci, et~al.
\newblock A spatiotemporal nonparametric {B}ayesian model of multi-subject
  f{MRI} data.
\newblock \emph{The Annals of Applied Statistics}, 10\penalty0 (2):\penalty0
  638--666, 2016.

\bibitem[Dunson(2009)]{dunson2009nonparametric}
David~B Dunson.
\newblock Nonparametric {B}ayes local partition models for random effects.
\newblock \emph{Biometrika}, 96\penalty0 (2):\penalty0 249--262, 2009.

\bibitem[Petrone et~al.(2009)Petrone, Guindani, and Gelfand]{petrone2009hybrid}
Sonia Petrone, Michele Guindani, and Alan~E Gelfand.
\newblock Hybrid {D}irichlet mixture models for functional data.
\newblock \emph{Journal of the {R}oyal {S}tatistical {S}ociety: {S}eries B
  ({S}tatistical {M}ethodology)}, 71\penalty0 (4):\penalty0 755--782, 2009.

\bibitem[Suarez and Ghosal(2016)]{suarez2016bayesian}
Adam~Justin Suarez and Subhashis Ghosal.
\newblock Bayesian clustering of functional data using local features.
\newblock \emph{Bayesian Analysis}, 11\penalty0 (1):\penalty0 71--98, 2016.

\bibitem[Margaritella et~al.(2021)Margaritella, In{\'a}cio, and
  King]{margaritella2021parameter}
Nicol{\`o} Margaritella, Vanda In{\'a}cio, and Ruth King.
\newblock Parameter clustering in {B}ayesian functional principal component
  analysis of neuroscientific data.
\newblock \emph{Statistics in Medicine}, 40\penalty0 (1):\penalty0 167--184,
  2021.

\bibitem[Wu and Li(2022)]{wu2022clustering}
Hui Wu and Yan-Fu Li.
\newblock Clustering spatially correlated functional data with multiple scalar
  covariates.
\newblock \emph{IEEE Transactions on Neural Networks and Learning Systems},
  2022.

\bibitem[Di et~al.(2009)Di, Crainiceanu, Caffo, and Punjabi]{di2009multilevel}
Chong-Zhi Di, Ciprian~M Crainiceanu, Brian~S Caffo, and Naresh~M Punjabi.
\newblock Multilevel functional principal component analysis.
\newblock \emph{The Annals of Applied Statistics}, 3\penalty0 (1):\penalty0
  458, 2009.

\bibitem[Serban and Jiang(2012)]{serban2012multilevel}
Nicoleta Serban and Huijing Jiang.
\newblock Multilevel functional clustering analysis.
\newblock \emph{Biometrics}, 68\penalty0 (3):\penalty0 805--814, 2012.

\bibitem[Richardson and Green(1997)]{richardson1997bayesian}
Sylvia Richardson and Peter~J Green.
\newblock On {B}ayesian analysis of mixtures with an unknown number of
  components (with discussion).
\newblock \emph{Journal of the {R}oyal {S}tatistical {S}ociety: {S}eries B
  ({S}tatistical {M}ethodology)}, 59\penalty0 (4):\penalty0 731--792, 1997.

\bibitem[Sethuraman(1994)]{sethuraman1994constructive}
Jayaram Sethuraman.
\newblock A constructive definition of {D}irichlet priors.
\newblock \emph{Statistica {S}inica}, 4\penalty0 (2):\penalty0 639--650, 1994.

\bibitem[Escobar(1994)]{escobar1994estimating}
Michael~D Escobar.
\newblock Estimating normal means with a {D}irichlet process prior.
\newblock \emph{Journal of the {A}merican {S}tatistical {A}ssociation},
  89\penalty0 (425):\penalty0 268--277, 1994.

\bibitem[Jara et~al.(2007)Jara, Garc{\'\i}a-Zattera, and
  Lesaffre]{jara2007dirichlet}
Alejandro Jara, Mar{\'\i}a~Jos{\'e} Garc{\'\i}a-Zattera, and Emmanuel Lesaffre.
\newblock A {D}irichlet process mixture model for the analysis of correlated
  binary responses.
\newblock \emph{Computational {S}tatistics \& {D}ata {A}nalysis}, 51\penalty0
  (11):\penalty0 5402--5415, 2007.

\bibitem[{de Valpine} et~al.(2017){de Valpine}, Turek, Paciorek,
  Anderson-Bergman, {Temple Lang}, and Bodik]{NIMBLE}
Perry {de Valpine}, Daniel Turek, Christopher Paciorek, Cliff Anderson-Bergman,
  Duncan {Temple Lang}, and Ras Bodik.
\newblock Programming with models: writing statistical algorithms for general
  model structures with {NIMBLE}.
\newblock \emph{Journal of Computational and Graphical Statistics},
  26:\penalty0 403--413, 2017.
\newblock \doi{10.1080/10618600.2016.1172487}.

\bibitem[{de Valpine} et~al.(2022){de Valpine}, Paciorek, Turek, Michaud,
  Anderson-Bergman, Obermeyer, {Wehrhahn Cortes}, Rodrìguez, {Temple Lang},
  and Paganin]{NIMBLEr}
Perry {de Valpine}, Christopher Paciorek, Daniel Turek, Nick Michaud, Cliff
  Anderson-Bergman, Fritz Obermeyer, Claudia {Wehrhahn Cortes}, Abel
  Rodrìguez, Duncan {Temple Lang}, and Sally Paganin.
\newblock \emph{{NIMBLE}: {MCMC}, Particle Filtering, and Programmable
  Hierarchical Modeling}, 2022.
\newblock URL \url{https://cran.r-project.org/package=nimble}.
\newblock {R} package version 0.12.2.

\bibitem[Krol et~al.(2018)Krol, Pawlitzki, Lotte, Gramann, and
  Zander]{krol2018sereega}
Laurens~R Krol, Juliane Pawlitzki, Fabien Lotte, Klaus Gramann, and Thorsten~O
  Zander.
\newblock Sereega: Simulating event-related eeg activity.
\newblock \emph{Journal of neuroscience methods}, 309:\penalty0 13--24, 2018.

\bibitem[Hubert and Arabie(1985)]{hubert1985comparing}
Lawrence Hubert and Phipps Arabie.
\newblock Comparing partitions.
\newblock \emph{Journal of {C}lassification}, 2\penalty0 (1):\penalty0
  193--218, 1985.

\bibitem[Wade and Ghahramani(2018)]{wade2018bayesian}
Sara Wade and Zoubin Ghahramani.
\newblock Bayesian cluster analysis: Point estimation and credible balls (with
  discussion).
\newblock \emph{Bayesian Analysis}, 13\penalty0 (2):\penalty0 559--626, 2018.

\bibitem[McDowell et~al.(2018)McDowell, Manandhar, Vockley, Schmid, Reddy, and
  Engelhardt]{mcdowell2018clustering}
Ian~C McDowell, Dinesh Manandhar, Christopher~M Vockley, Amy~K Schmid,
  Timothy~E Reddy, and Barbara~E Engelhardt.
\newblock Clustering gene expression time series data using an infinite
  {G}aussian process mixture model.
\newblock \emph{PLoS {C}omputational {B}iology}, 14\penalty0 (1):\penalty0
  e1005896, 2018.

\bibitem[Zhang et~al.(1995)Zhang, Begleiter, Porjesz, Wang, and
  Litke]{zhang1995event}
Xiao~Lei Zhang, Henri Begleiter, Bernice Porjesz, Wenyu Wang, and Ann Litke.
\newblock Event related potentials during object recognition tasks.
\newblock \emph{Brain Research Bulletin}, 38\penalty0 (6):\penalty0 531--538,
  1995.

\bibitem[Yan et~al.(2023)Yan, Ding, Li, Wu, and Zhu]{yan2023effect}
Chunping Yan, Qianqian Ding, Yunyun Li, Meng Wu, and Jinfu Zhu.
\newblock Effect of retrieval reward on episodic recognition with different
  difficulty: {ERP} evidence.
\newblock \emph{International Journal of Psychophysiology}, 183:\penalty0
  41--52, 2023.

\bibitem[Saltzmann et~al.(2022)Saltzmann, Moen, Chaisson, Eich, Fan, Beck, and
  Lucas]{saltzmann2022neural}
Stephanie Saltzmann, Katherine Moen, Felicia Chaisson, Brandon Eich, Gaojie
  Fan, Melissa Beck, and Heather Lucas.
\newblock Neural correlates of task-irrelevant feature processing in visual
  working memory.
\newblock \emph{Journal of Vision}, 22\penalty0 (14):\penalty0 3567--3567,
  2022.

\bibitem[Stevens et~al.(2023)Stevens, Teich, Longenecker, and
  Sponheim]{stevens4208065relational}
Kara~L Stevens, Collin~D Teich, Julia~M Longenecker, and Scott~R Sponheim.
\newblock Relational memory function in schizophrenia: Electrophysiological
  evidence for early perceptual and late associative abnormalities.
\newblock \emph{Schizophrenia Research}, 254:\penalty0 99--108, 2023.

\bibitem[Sur and Sinha(2009)]{sur2009event}
Shravani Sur and VK~Sinha.
\newblock Event-related potential: An overview.
\newblock \emph{Industrial {P}sychiatry {J}ournal}, 18\penalty0 (1):\penalty0
  70, 2009.

\bibitem[Snodgrass and Vanderwart(1980)]{snodgrass1980standardized}
Joan~G Snodgrass and Mary Vanderwart.
\newblock A standardized set of 260 pictures: norms for name agreement, image
  agreement, familiarity, and visual complexity.
\newblock \emph{Journal of {E}xperimental {P}sychology: {H}uman {L}earning and
  {M}emory}, 6\penalty0 (2):\penalty0 174, 1980.

\bibitem[Crossman and Neary(2018)]{crossman2018neuroanatomy}
Alan~R Crossman and David Neary.
\newblock \emph{Neuroanatomy E-book: An Illustrated Colour Text}.
\newblock Elsevier Health Sciences, 2018.

\bibitem[Ramsay et~al.(2020)Ramsay, Graves, and Hooker]{fdapackage}
J.~O. Ramsay, Spencer Graves, and Giles Hooker.
\newblock \emph{fda: Functional Data Analysis}, 2020.
\newblock URL \url{https://CRAN.R-project.org/package=fda}.
\newblock R package version 5.1.4.

\bibitem[Begleiter et~al.(1993)Begleiter, Porjesz, and
  Wang]{begleiter1993neurophysiologic}
H~Begleiter, B~Porjesz, and W~Wang.
\newblock A neurophysiologic correlate of visual short-term memory in humans.
\newblock \emph{Electroencephalography and Clinical Neurophysiology},
  87\penalty0 (1):\penalty0 46--53, 1993.

\bibitem[Begleiter et~al.(1995)Begleiter, Porjesz, and
  Wang]{begleiter1995event}
Henri Begleiter, Bernice Porjesz, and Wenyu Wang.
\newblock Event-related brain potentials differentiate priming and recognition
  to familiar and unfamiliar faces.
\newblock \emph{Electroencephalography and Clinical Neurophysiology},
  94\penalty0 (1):\penalty0 41--49, 1995.

\bibitem[Loader(2006)]{loader2006local}
Clive Loader.
\newblock \emph{Local Regression and Likelihood}.
\newblock Springer Science \& Business Media, 2006.

\bibitem[Grzesik(2017)]{grzesik2017local}
Katherine Grzesik.
\newblock \emph{Local Cross-Validated Smoothing Parameter Estimation for Linear
  Smoothers}.
\newblock PhD thesis, University of Rochester, 2017.

\bibitem[Dimsdale-Zucker et~al.(2022)Dimsdale-Zucker, Maciejewska, Kim,
  Yonelinas, and Ranganath]{Dimsdalezucker2022108287}
Halle~R. Dimsdale-Zucker, Karina Maciejewska, Kamin Kim, Andrew~P. Yonelinas,
  and Charan Ranganath.
\newblock Individual differences in behavioral and electrophysiological
  signatures of familiarity- and recollection-based recognition memory.
\newblock \emph{Neuropsychologia}, 173:\penalty0 108287, 2022.

\bibitem[Turano et~al.(2016)Turano, Marzi, and Viggiano]{Turano20161}
Maria~Teresa Turano, Tessa Marzi, and Maria~Pia Viggiano.
\newblock Individual differences in face processing captured by {ERP}s.
\newblock \emph{International Journal of Psychophysiology}, 101:\penalty0 1--8,
  2016.

\bibitem[Laszlo and Sacchi(2015)]{LaszloO201542}
Sarah Laszlo and Elizabeth Sacchi.
\newblock Individual differences in involvement of the visual object
  recognition system during visual word recognition.
\newblock \emph{Brain and Language}, 145-146:\penalty0 42--52, 2015.

\bibitem[Mouraux and Iannetti(2008)]{mouraux2008across}
Andr{\'e} Mouraux and Gian~Domenico Iannetti.
\newblock Across-trial averaging of event-related eeg responses and beyond.
\newblock \emph{Magnetic resonance imaging}, 26\penalty0 (7):\penalty0
  1041--1054, 2008.

\bibitem[Cozzini et~al.(2014)Cozzini, Jasra, Montana, and
  Persing]{cozzini2014bayesian}
Alberto Cozzini, Ajay Jasra, Giovanni Montana, and Adam Persing.
\newblock A bayesian mixture of lasso regressions with t-errors.
\newblock \emph{Computational Statistics \& Data Analysis}, 77:\penalty0
  84--97, 2014.

\bibitem[Shi et~al.(2022)Shi, Yang, Wang, Ma, Beg, Pei, and Cao]{shi2022two}
Haolun Shi, Yuping Yang, Liangliang Wang, Da~Ma, Mirza~Faisal Beg, Jian Pei,
  and Jiguo Cao.
\newblock Two-dimensional functional principal component analysis for image
  feature extraction.
\newblock \emph{Journal of Computational and Graphical Statistics}, 31\penalty0
  (4):\penalty0 1127--1140, 2022.

\bibitem[Happ and Greven(2018)]{happ2018multivariate}
Clara Happ and Sonja Greven.
\newblock Multivariate functional principal component analysis for data
  observed on different (dimensional) domains.
\newblock \emph{Journal of the American Statistical Association}, 113\penalty0
  (522):\penalty0 649--659, 2018.

\bibitem[Rodriguez et~al.(2008)Rodriguez, Dunson, and
  Gelfand]{rodriguez2008nested}
Abel Rodriguez, David~B Dunson, and Alan~E Gelfand.
\newblock The nested {D}irichlet process.
\newblock \emph{Journal of the American statistical Association}, 103\penalty0
  (483):\penalty0 1131--1154, 2008.

\bibitem[Teh et~al.(2006)Teh, Jordan, and Beal]{teh2006w}
YW~Teh, MI~Jordan, and MJ~Beal.
\newblock Hierarchical dirichlet processes.
\newblock \emph{Journal of the American Statistical Association}, 101\penalty0
  (476):\penalty0 1566--1581, 2006.

\bibitem[Wulsin et~al.(2012)Wulsin, Jensen, and Litt]{wulsin2012hierarchical}
Drausin Wulsin, Shane Jensen, and Brian Litt.
\newblock A hierarchical {D}irichlet process model with multiple levels of
  clustering for human eeg seizure modeling.
\newblock \emph{arXiv preprint arXiv:1206.4616}, 2012.

\bibitem[King et~al.(2023)King, Sarzo, and Elvira]{king2022large}
Ruth King, Blanca Sarzo, and V{\'\i}ctor Elvira.
\newblock When ecological individual heterogeneity models and large data
  collide: an importance sampling approach.
\newblock \emph{Annals of Applied Statistics}, 17\penalty0 (4):\penalty0
  3112--3132, 2023.

\bibitem[Su(2020)]{su2020divide}
Ya~Su.
\newblock A divide and conquer algorithm of {B}ayesian density estimation.
\newblock \emph{arXiv preprint arXiv:2002.07094}, 2020.

\bibitem[Song et~al.(2020)Song, Wang, and Dunson]{song2020distributed}
Hanyu Song, Yingjian Wang, and David~B Dunson.
\newblock Distributed {B}ayesian clustering using finite mixture of mixtures.
\newblock \emph{arXiv preprint arXiv:2003.13936}, 2020.

\bibitem[Quiroz et~al.(2018)Quiroz, Kohn, Villani, and
  Tran]{quiroz2018speeding}
Matias Quiroz, Robert Kohn, Mattias Villani, and Minh-Ngoc Tran.
\newblock Speeding up {MCMC} by efficient data subsampling.
\newblock \emph{Journal of the American Statistical Association}, 2018.

\end{thebibliography}
\newpage
\phantom{aaaa}
\end{document}

\maketitle
%\newpage
\section{SUPPLEMENTARY MATERIALS A:\\ Hyperparameter values}
Hyperparameters setting for the simulation and ERPs studies of Sections 3 and 4.
\begin{table}[h!]
	\small\sf\centering
	\begin{tabular}{lc|ll|}
		\multicolumn{2}{l|}{\textbf{Parameter}} & \multicolumn{2}{l|}{\textbf{Empirical estimate/ value}} \\ \hline
		&$h_{k1}^{-1}$ & $2\times$ Variance of the bootstrap means of all empirical fPC scores.&            
		\vspace{2pt}    \\ 
		&$\gamma_{k1}$ &$(\text{standard deviation of the empirical fPC scores})^{2}$. &            
		\vspace{2pt}    \\  
		&$\phi_{kD}$ & mean of empirical fPC scores of group $\text{D}=\left\{1,2\right\}$. &            
		\vspace{2pt}    \\  
		&$h_{kD}^{-1}$ & $2\times$ Variance of bootstrap means of 50\% empirical fPC scores of group $\text{D}=\left\{1,2\right\}$. &            
		\vspace{2pt}    \\ 
		&$\gamma_{kD}$ & $(\text{standard deviation of the empirical fPC scores of group D}=\left\{1,2\right\})^{2}$. &            
		\vspace{2pt}    \\ 
		&$h_{ukj}^{-1}$ & square of $(1/2.5)\times$ range of the empirical fPC scores of group $\text{D}=\left\{1,2\right\}$. &            
		\vspace{2pt}    \\ 
		&$\gamma_{ukj}$ & $(\text{standard deviation of the empirical fPC scores of group D}=\left\{1,2\right\})^{2}$. &            
		\vspace{2pt}    \\
		&$\boldsymbol{\delta}$ & $c(9/20,9/20,2/20)$ &            
		\vspace{2pt}    \\
		&$\alpha_{k}$ & 1 &            
		\vspace{2pt}    \\ \hline

	\end{tabular}
	\caption{Hyperparameters setting for the simulation and ERPs studies}
\end{table}
\section{SUPPLEMENTARY MATERIALS B:\\ Joint posterior distribution}
The joint posterior distribution of all parameters $\boldsymbol{\Theta}=(\boldsymbol{\xi},\boldsymbol{g}, \boldsymbol{\eta},\boldsymbol{\mu}, \boldsymbol{s},\boldsymbol{\omega},\tau)$ can be written as follows:
\begin{align*}
	\pi(\boldsymbol{\Theta}|\mathbf{\tilde{Y}}, \boldsymbol{\nu},\boldsymbol{\phi},\boldsymbol{h}, \boldsymbol{\gamma},\boldsymbol{\delta}, \boldsymbol{\alpha})&\propto \tau^{UnT/2}\exp\bigg\{-\frac{\tau}{2}\sum_{u=1}^{U}\sum_{i=1}^{n}\sum_{t=1}^{T}\Big(\tilde{Y}_{uit}-\sum_{k=1}^{K}\xi_{uik}\nu_{kt}\Big)^{2}\bigg\}\\\nonumber
	&\quad \bigg\{\prod_{u=1}^{U}\prod_{k=1}^{K}\prod_{j=1}^{J}s_{ukj}^{1/2}\bigg\}\exp\Big\{-\frac{1}{2}\sum_{u=1}^{U}\sum_{k=1}^{K}\sum_{j=1}^{J}s_{ukj}(\xi_{ukj}-\mu_{ukj})^{2}\Big\}\\\nonumber
	&\quad
	\bigg\{\prod_{u=1}^{U}\prod_{k=1}^{K}\prod_{\Omega=1}^{3}\omega_{\Omega k}^{[g_{uk}=\Omega]}\bigg\}\bigg\{\prod_{k=1}^{K}\prod_{D=1}^{2}\prod_{j=1}^{J}p_{jDk}^{n_jDk}\bigg\}\\\nonumber
	&\quad
	\bigg\{\prod_{k=1}^{K}\prod_{D=1}^{2}\prod_{j=1}^{J}\binom{n_{Dk}}{n_{jDk}}p_{jDk}^{*^{n_{jDk}}}(1-p_{jDk}^{*})^{n_{Dk}-n_{jDk}}\bigg\}\bigg\{\prod_{k=1}^{K}\prod_{D=1}^{2}\prod_{j=1}^{J}\alpha_{k}(1-p_{jDk}^{*})^{\alpha_{k}-1}\bigg\}\\\nonumber
	%\bigg\{ \prod_{k=1}^{K}\dfrac{\Gamma(\alpha_{k})}{\Gamma(n-\alpha_{k})}\bigg[\prod_{j=1}^{J}\dfrac{\Gamma(n_{jDk}+\alpha_{k}/J)}{\Gamma(\alpha_{k}/J)}\bigg]\bigg\}\\\nonumber
	&\quad
	\bigg[\dfrac{\Gamma(\sum_{\Omega=1}^{3}\delta_{\Omega})}{\prod_{\Omega=1}^{3}\Gamma(\delta_{\Omega})}\bigg]^{K}\bigg\{\prod_{k=1}^{K}\prod_{\Omega=1}^{3}\omega_{\Omega k}^{\delta_{\Omega}-1}\bigg\} \bigg\{\prod_{k=1}^{K}h_{k1}^{1/2}\bigg\}\exp\Big\{-\frac{1}{2}\sum_{k=1}^{K}h_{k1}(\mu_{k1})^{2}\Big\}\\\nonumber
	&\quad
	\bigg\{\prod_{k=1}^{K}\prod_{D=1}^{2}h_{kD}^{1/2}\bigg\}\exp\Big\{-\frac{1}{2}\sum_{k=1}^{K}\sum_{D=1}^{2}h_{kD}(\mu_{kD}-\phi_{kD})^{2}\Big\}\\\nonumber
	&\quad
	\bigg\{\prod_{u=1}^{U}\prod_{k=1}^{K}\prod_{j=1}^{J}h_{ukj}^{1/2}\bigg\}\exp\Big\{-\frac{1}{2}\sum_{u=1}^{U}\sum_{k=1}^{K}\sum_{j=1}^{J}h_{ukj}(\mu_{ukj})^{2}\Big\}\\\nonumber
	&\quad
	\bigg\{\prod_{k=1}^{K}\dfrac{1}{\gamma_{k1}}\bigg\}\bigg\{\prod_{k=1}^{K}\prod_{D=1}^{2}\dfrac{1}{\gamma_{kD}}\bigg\}\bigg\{\prod_{u=1}^{U}\prod_{k=1}^{K}\prod_{j=1}^{J}\dfrac{1}{\gamma_{ukj}}\bigg\}\tau^{a-1}\exp\bigg\{-b\tau\bigg\},
\end{align*}
where  $[g_uk=\Omega]$ is equal to 1 when $g_uk=\Omega$ and zero otherwise; $n_{jDk}$ is the number of fPC scores in group D and dimension k that are assigned to the $j$th cluster; $n_{Dk}=\sum_{j=1}^{J}n_{jDk}$ and  $a=b=10^{-3}$.  A list of the hyperparameters values used in the simulation and ERPs studies is given in Supplementary Materials A.

\newpage
\section{SUPPLEMENTARY MATERIALS C:\\ Sensitivity Analysis}
Table 2 reports the different scenarios tested in our prior sensitivity analysis. The partition in the first eigendimenson was robust to changes in the hyperparameter values. The partition in the second eigendimension was also retrieved in all the scenarios tested; however, some MCMC chains converged to slightly different partitions with group clusters including between $ 25\% $ and $45\%$ of the subjects' fPC scores. We note that the second dimension had the highest heterogeneity in all scenarios tested, with subject-specific clusters always allocated to $30\% - 35\%$ of the subjects' fPC scores. 
\begin{table}[h!]
	\small\sf\centering
	\begin{tabular}{c|ll}
		\multicolumn{3}{c}{\textbf{Scenario 1}} \\
		\multicolumn{1}{l|}{\textbf{Parameter}} & \multicolumn{2}{l}{\textbf{Empirical estimate/ value}}\vspace{2pt}\\ \hline 
		$\boldsymbol{\gamma_{k1}}$ & $(\text{standard deviation of the empirical fPC scores})^{2.2}$. &            
		\vspace{2pt}    \\  
		$\boldsymbol{\gamma_{kD}}$ & $(\text{standard deviation of the empirical fPC scores of group D} \left\{1,2\right\})^{2.2}$. &            
		\vspace{2pt} \\ 
		$\boldsymbol{\gamma_{ukj}}$ & $(\text{standard deviation of the empirical fPC scores of group D}=\left\{1,2\right\})^{2.2}$. &                  
		\vspace{2pt}    \\ \hline
	\multicolumn{3}{c}{\textbf{Scenario 2}} \\ 
		$\boldsymbol{h_{k1}^{-1}}$ & $4\times$ Variance of the bootstrap means of all empirical fPC scores.&                      
\vspace{2pt}    \\  
$\boldsymbol{h_{kD}^{-1}}$ & $4\times$ Variance of bootstrap means of 50\% empirical fPC scores of group $\text{D}=\left\{1,2\right\}$. &            
\vspace{2pt}    \\	 \hline
	\multicolumn{3}{c}{\textbf{Scenario 3}} \\ 
	$\boldsymbol{\delta}$ & $c(4/10,4/10,2/10)$ &                     
	\vspace{2pt}    \\ \hline
	\multicolumn{3}{c}{\textbf{Scenario 4}} \\
	$\boldsymbol{\delta}$ & $c(1/3,1/3,1/3)$ &                    
	\vspace{2pt}    \\ \hline
\multicolumn{3}{c}{\textbf{Scenario 5}} \\
$\boldsymbol{\alpha_{k}}$ & $0.5$ &                    
\vspace{2pt}    \\ \hline
\multicolumn{3}{c}{\textbf{Scenario 6}} \\
$\boldsymbol{\alpha_{k}}$ & $2$ &                    
\vspace{2pt}    \\ \hline
\end{tabular}
\caption{ERPs study - sensitivity analysis: each scenario highlights the different hyperparameter values tested. The estimated partitions discussed in the paper were retrieved under all the scenarios tested.}
\end{table}
\newpage
\section{SUPPLEMENTARY MATERIALS D:\\ ERPs recordings, 3rd eigendimension and fPCAs at each location}
The ERPs recorded from ten subjects under the two  experimental conditions (matched and unmatched visual stimuli) are plotted in Figure 1 while Figures 2 and 3 show the third eigendimension and the relative clustering of the functional PC scores obtained with our model, respectively. This eigendimension was excluded from the final interpretation as it represented mostly the variation arising from a single individual.
\begin{figure}[!h]
	\centering
	\includegraphics[width=0.95\linewidth]{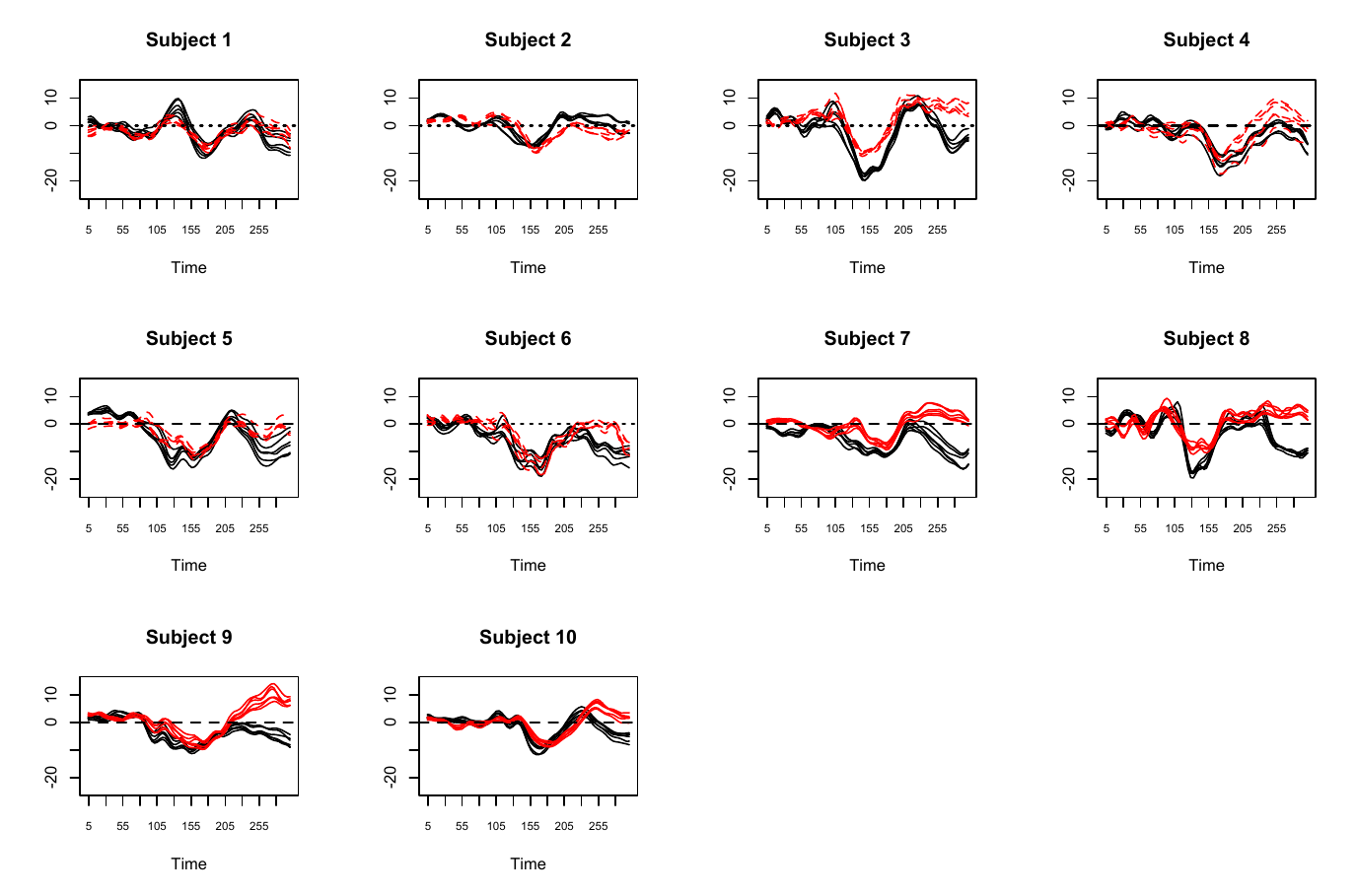}
	\caption{ERPs study: recordings of all subjects included in the analysis. ERPs recorded during matched stimuli are in black (solid), those recorded during unmatched stimuli are in red (dashed).}
	\label{FigApC1} 
\end{figure}
\begin{figure}[!h]
	\centering
	\includegraphics[width=0.88\linewidth]{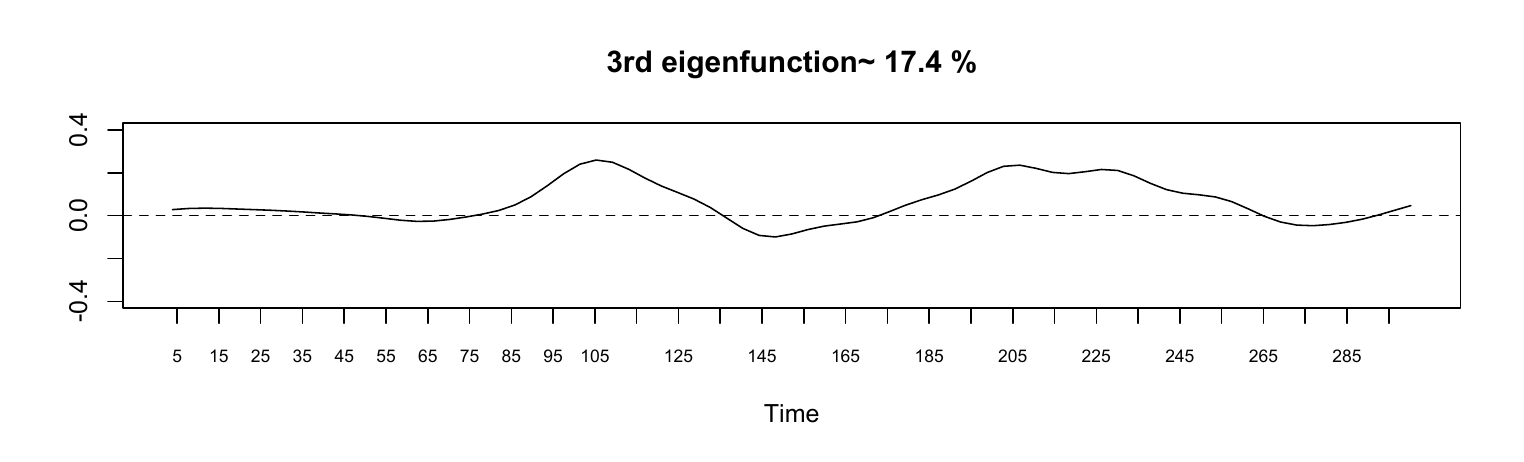}
	\caption{ERPs study: third eigendimension.}
	\label{FigApC2} 
\end{figure}
\begin{figure}[!h]
	\centering
	\includegraphics[width=0.88\linewidth]{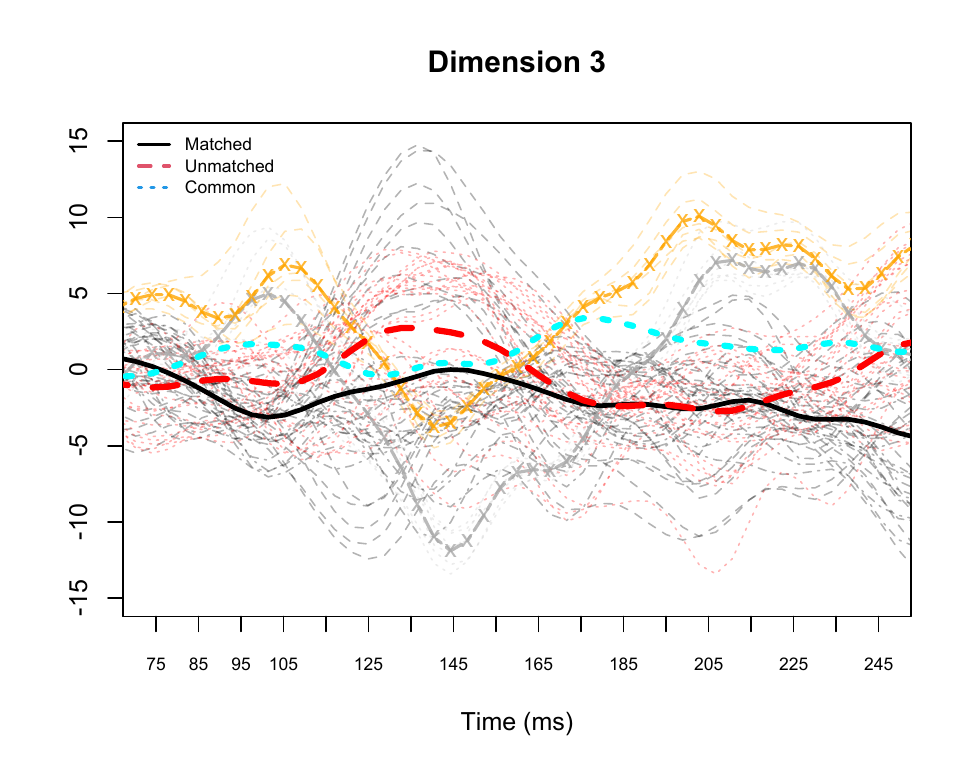}
	\caption{ERPs study: local group differences between matched (solid) and unmatched (dashed) recordings identified by the model in the third eigendimension. In bold the common and group cluster means are superimposed on the relative recordings. The symbols x (grey) and '$+$' (orange) identify individual-specific cluster means. }
	\label{FigApC3} 
\end{figure}
\clearpage
Results of fPCAs carried out at each electrode show minor spatial variation in the shape of the trajectories recorded from different subjects. This was expected, as the electrodes considered were all recording from the occipital area of the brain under the same experimental conditions. 
The location-specific eigenvalues range  $[0.334,0.390]$, $[0.173, 0.319]$, and, $[0.093, 0.168]$ for FPCA 1 to 3, respectively. Figure 4 and 5 shows the distribution of the eigenfunctions and eigenvalues in the three eigendimensions.
\begin{figure}[h!]
	\centering
	\includegraphics[width=0.99\linewidth]{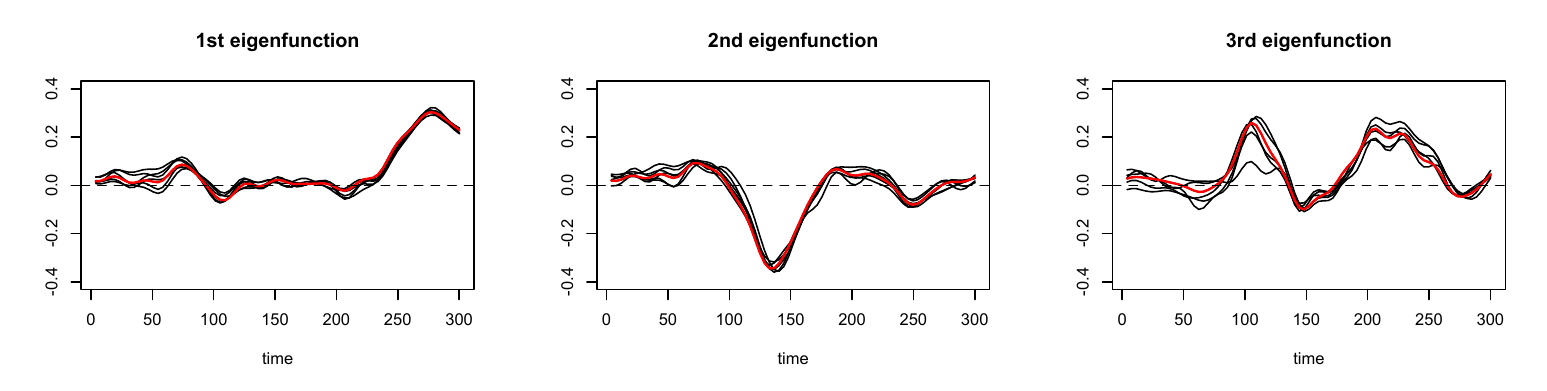}
	\caption{Case study: the three leading eigenfunction computed for all locations (red) and for each location separately (black).}
	\label{Fig1R} 
\end{figure}
\begin{figure}[h!]
	\centering
	\includegraphics[width=0.99\linewidth]{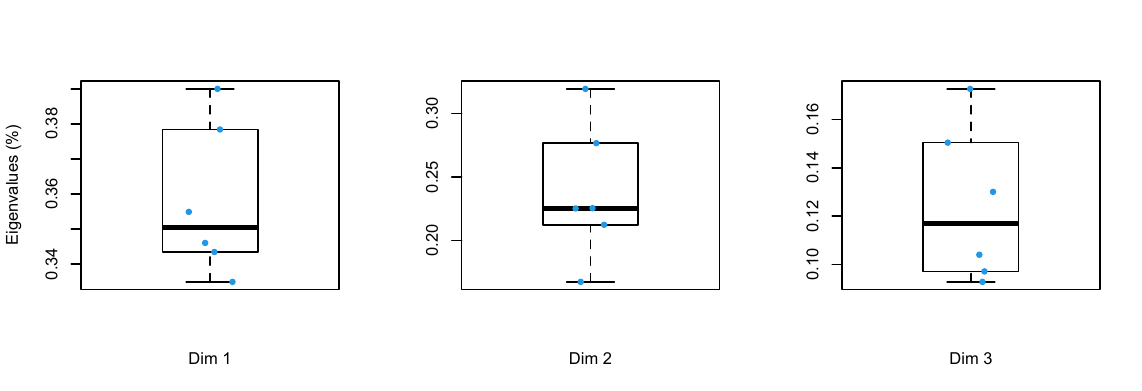}
	\caption{Case study: Distribution of the location-specific eigenvalues in the three leading eigenfunctions.}
	\label{Fig2R} 
\end{figure}

\clearpage
\section{SUPPLEMENTARY MATERIALS E:\\ MCMC checks} 
\begin{figure}[!h]
	\centering
	\includegraphics[width=0.92\linewidth]{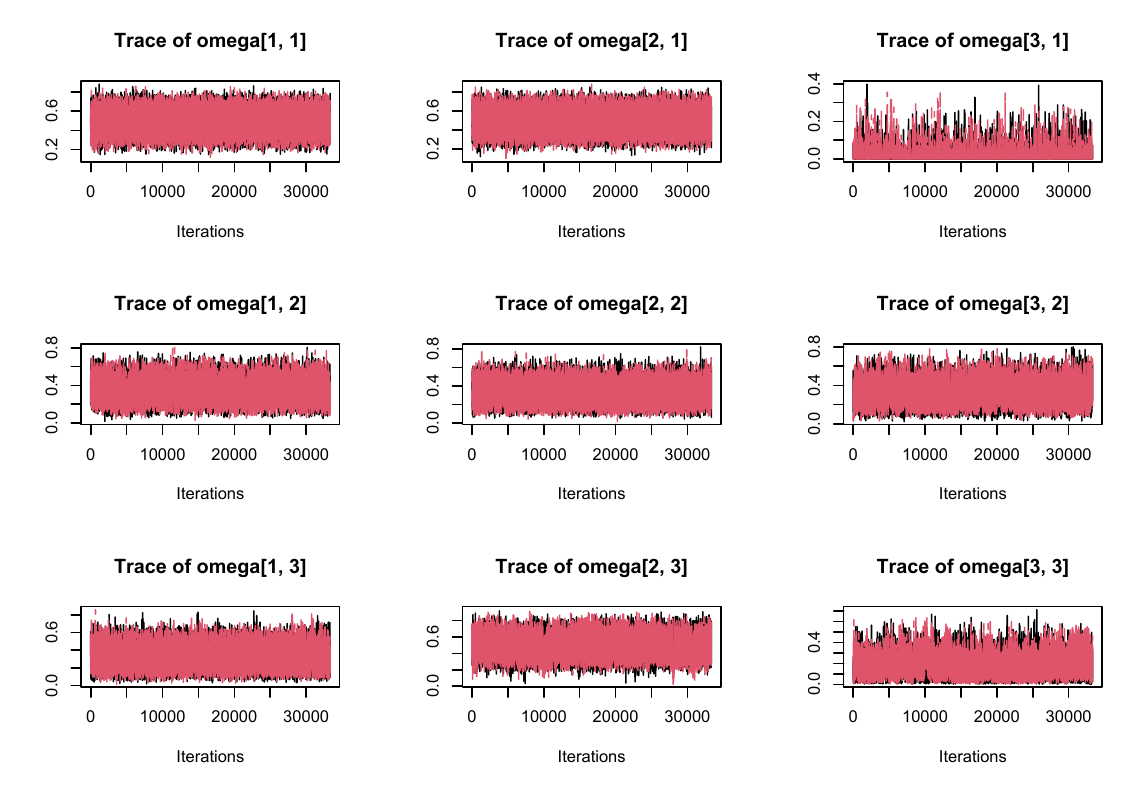}
	\caption{ERPs study: MCMC checks. A sample of traceplots supporting convergence of the main parameters of our models. Here are shown the MCMC chains of the $\omega$ parameters in all three eigendimensions.}
	\label{FigApD1} 
\end{figure}

\begin{figure}[!h]
	\centering
	\includegraphics[width=0.92\linewidth]{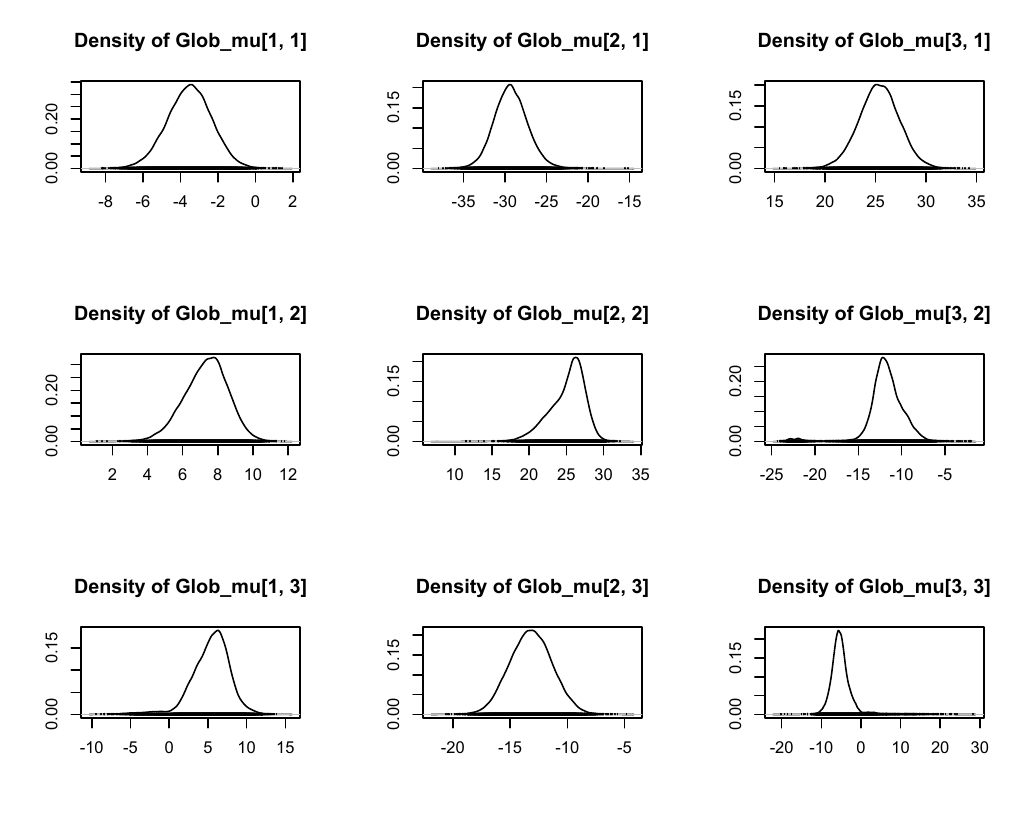}
	\caption{ERPs study: MCMC checks. An example of density plots showing unimodal posterior densities for the Common ([1, k]) and group-specific ([2, k] and [3, k]) mean  parameters in all three k eigendimensions.}
	\label{FigApD2} 
\end{figure}

Convergence checks were carried out both visually and quantitatively, using traceplots and the Brooks-Gelman-Rubin statistics. Traceplots and density plots showed overlapping chains, fast exploration of the parameter space and unimodal posteriors (Figures~\ref{FigApD1} and \ref{FigApD2} show examples of traceplots and density plots, respectively). The BGR statistics of 93\% of the parameters of interest were below the 1.1 threshold. The five parameters with values above the threshold (the maximum was 1.6) had no trend and good overall overlapping of the chains and therefore no further action was taken. Chain length was monitored using the effective sample size (ESS). The ESS of 96\% of the parameters of interest were all above $10^3$. Of the only three parameters with ESS below 1000, the minimum was 350, which was considered a sufficient number of effective samples to reliably estimate a posterior mean. 
